\documentclass[twocolumn]{biophys-new}
\usepackage[utf8]{inputenc}
\usepackage{graphicx}
\usepackage[colorlinks,allcolors=cyan!70!black]{hyperref}

\usepackage{lipsum}
\usepackage{amsmath}

\title{Nanopillar-Driven Antibacterial Surfaces: Elucidating Bactericidal Mechanisms and Engineering Nanostructures for Enhanced Efficacy}
\runningtitle{Acta Biomaterialia} %% For page header

\author[1]{Akash Singh}
\author[2]{Yi Zhang}
\author[3]{Qing Cao}
\author[4,*]{Yumeng Li}
\runningauthor{Akash Singh, Yi Zhang, Qing Cao and Yumeng Li} %% For page header

\affil[1]{University of Illinois at Urbana Champaign, 104 S Mathews Ave Urbana, IL 61801}
\affil[2]{University of Illinois at Urbana Champaign, 104 1008 Superconductivity Center Urbana, IL 61801}
\affil[3]{University of Illinois at Urbana Champaign, 104 1008 Superconductivity Center Urbana, IL 61801}
\affil[4]{University of Illinois at Urbana Champaign, 104 S Mathews Ave Urbana, IL 61801}

\corrauthor[*]{yumengl@ilinois.edu}

\papertype{Article}
\begin{document}

\begin{frontmatter}

\begin{abstract}
Insects, such as dragonflies and cicadas, have evolved nanoprotusions on wing surfaces that can rupture bacterial membranes upon contact. This remarkable natural phenomenon has inspired researchers worldwide to develop antibacterial implant surfaces using synthetic materials that mimic insect wings. To design these biomimetic bactericidal surfaces effectively, it is crucial to understand how nanopillars mechanically interact with bacteria and induce bactericidal effects. However, a significant challenge in studying this interaction lies in the small length and time scales associated with bactericidal actions.
Molecular Dynamics (MD) simulations are well established for accurately modeling materials at small scales, offering high precision with relatively low computational costs. In this study, we developed a coarse-grained membrane model to investigate the dynamic mechanical responses of gram-positive and gram-negative bacterial membranes under various nanopillar array configurations. We also explored the bactericidal mechanisms associated with bacteria of two geometric shapes, i.e. spherical and cylindrical, by varying the bacterial membrane bending rigidity and the loading rate which is the acceleration at which bacteria impinge upon the nanopillared substrate.
Our findings reveal two distinct failure mechanisms based on the bending rigidity of bacterial membranes upon contact with nanopillared surfaces. For membranes with low membrane bending rigidity, which mimic gram-negative bacteria, tearing occurs near the nanopillar tip rather than in the sagged region between two nanopillars, contrary to what has been proposed as the primary bactericidal mechanism in previous studies. Conversely, for membranes with high bending rigidity, which mimic gram-positive bacteria, the primary mode of failure is piercing or puncturing at the point of contact between the nanopillar and the bacterial membrane.
Moreover, gram-positive bacteria with high membrane bending rigidity are less likely to be killed under the same nanopillar configuration and loading rate that is lethal for gram-negative bacteria with low membrane bending rigidity. A threefold increase in the loading rate is required for the piercing mechanism to activate and effectively kill gram-positive bacteria.
Additionally, we investigated the effects of nanopillar spacing and height on bactericidal activity. Our results indicate that these parameters significantly influence bactericidal efficacy. Specifically, increasing the nanopillar height beyond a critical threshold compared to the geometric height of the bacteria substantially enhances bactericidal activity, whereas increasing the nanopillar spacing beyond the geometric width of the bacteria significantly reduces it. This research introduces a simplified, single-layered coarse-grained model of bacterial membranes that captures bending rigidity with high fidelity. This new model enables the simulation of dynamic bactericidal activity for entire gram-negative and gram-positive bacterial cells at a fraction of the computational cost compared to all-atom models used in previous studies. With this model, it is possible to simulate full-scale bacterial interactions over extended timeframes, allowing for a more realistic representation of bactericidal activity. By offering valuable insights into the precise design of nanopillared surfaces—including critical parameters such as nanopillar height and spacing—this study provides a foundation for optimizing bactericidal efficacy. Ultimately, these findings contribute to the advancement of chemical-free antibacterial surfaces.
\end{abstract}

\begin{sigstatement}
Bactericidal effects in nanopillared surfaces, Coarse grained model of bacteria, Molecular dynamics simulation, Bio-mimetic surfaces, Mechano-bactericidal mechanism, Nanostructured surfaces
\end{sigstatement}
\end{frontmatter}

\section*{Introduction}

It is widely recognized that certain insect wings, such as those of cicadas and dragonflies, exhibit remarkable antibacterial and antifungal properties. Previous research has revealed a fascinating bactericidal mechanism in which the physical nanoprotrusions on an insect's wing surface stretch and damage microbial cells upon contact, ultimately causing their rupture and demise \cite{Ivanova2012_1,Ivanova2013_2}.
This phenomenon was first documented in Pseudomonas aeruginosa (P. aeruginosa) by Ivanova et al. \cite{Ivanova2012_1}. They observed individual bacteria navigating between the nanopillars on cicada wings, with their survival limited to cycles as short as 20 minutes. The mechanical rupture of the bacterial outer membrane, facilitated by the nanopillared substrate, was proposed as the primary cause of their demise. Notably, these intriguing bactericidal effects are not limited to P. aeruginosa; they also extend to other gram-negative bacteria such as Escherichia coli (E. coli), Branhamella catarrhalis, and Pseudomonas fluorescens \cite{Hasan2013_6}. Beyond cicadas, the wings of dragonflies (Diplacodes bipunctata) exhibit potent bactericidal properties. These wings can kill both gram-positive bacteria, such as Staphylococcus aureus (S. aureus), and gram-negative bacteria, including Bacillus subtilis and P. aeruginosa \cite{Ivanova2013_2}. The unique architecture of Diplacodes bipunctata wings, characterized by their capillary-like structures, is believed to amplify stress and deformation within bacterial membranes, enhancing their bactericidal efficacy even against gram-positive cells \cite{Ivanova2013_2}. The bactericidal properties of cicada and dragonfly wings have captivated researchers over the past decade due to their ability to physically kill bacteria. This capability presents a promising strategy for addressing various challenges, such as preventing biofilm formation and minimizing infection risks associated with implantable medical devices in humans and animals. Currently, most implantable medical devices rely on materials infused with antibiotics to achieve bactericidal effects, which primarily kill bacteria through biochemical processes. However, this approach has significant limitations, such as the need to frequently remove and reapply antibiotic coatings, potential side effects on human cells, and the unintended elimination of essential non-target bacteria in the human body. To address these issues, researchers have explored various nanofabrication techniques to create bactericidal nanotopographies on synthetic materials. These techniques have been applied to a wide range of materials, including black silicon (bSi) \cite{Ivanova2013_2}, titanium metal \cite{Diu2014_7}, titanium alloys \cite{Terje2016_8}, and various polymers \cite{Hazell2018_14}. By developing these innovative nanotopographies, researchers aim to revolutionize antibacterial strategies and medical implant technologies.

Over the years, various theoretical models have been developed to elucidate the intricate process of bacterial contact killing by nanopillared substrates. One such model, known as the biophysical model, postulates that the nanopillar arrays on cicada wings physically stretch the bacterial cell membrane upon contact, ultimately leading to the rupture and death of the bacterial cell. The bending rigidity of the bacterial cell membrane plays a critical role in determining its susceptibility to failure, as highlighted by the study of Pogodin et al. \cite{Pogodin2013_10}.
In support of this concept, Pogodin et al. proposed an elastic mechanical model suggesting that gram-positive bacteria exhibit greater resistance to deformation and rupture by nanopillars due to their lower maximum stretching capacity and higher membrane bending rigidity. However, this model faced limitations in accurately approximating the curvature of bacteria while evaluating the adsorption of the bacterial membrane onto nanopillared surfaces. Another theoretical model by Xue et al. \cite{XUE2015_15} proposed that, given the bending rigidity of the bacterial cell wall or outer membrane, the bactericidal efficiency of nanopillared surfaces is influenced by the geometric parameters of the surface structures. The study suggested that enhanced antibacterial properties could be achieved by decreasing the nanopillar diameter and increasing the nanopillar spacing. Additionally, a quantitative thermodynamic model by Xinelei et al. \cite{Xinlei_16} proposed that the bactericidal activity of nanopillared surfaces is intricately linked to the balance between deformation and adhesion energies of the nanopillars and the bacterial outer membrane. This model predicts that within specific dimensions, such as nanopillar radii between 0 and 50 nm and nanopillar spacing ranging from 100 to 250 nm, the bacterial outer membrane experiences maximum stretching, resulting in the highest bactericidal efficiency. Watson et al. \cite{Watson2019_Asimple} introduced another theoretical model based on surface energies to explain the physical mechanism of bacterial lysis by nanopillared surfaces. Watson et al. proposed that the bacterial cell wall ruptures only when the tensile stress in the membrane caused by stretching exceeds the tensile strength of the cell wall. Although all these biophysical models have contributed to understanding the bactericidal nature of nanopillared surfaces, they still fall short in certain areas. Specifically, they cannot fully reveal the exact failure modes of the bacterial membrane, provide precise quantitative evaluations of bactericidal efficiency across various nanostructured surfaces, or consistently align with experimental findings.

Numerous experimental studies have identified various factors influencing the antimicrobial effectiveness of nanopillared surfaces. One critical factor is the adhesive strength between bacteria and nanopillars, which directly impacts the efficiency of bacterial elimination. Research by Nowlin et al. \cite{Nowlin2014_9} has shown that the highest strain and rupture occur in bacteria, such as Saccharomyces cerevisiae, that strongly adhere to cicada wing nanopillars. Bandara et al. \cite{Bandara2017_4} suggested that the jagged geometry of nanopillared arrays contributes to the bactericidal effects observed in the wings of certain insects. Initially, bacteria attach to a few relatively taller pillars via the secretion of extracellular polymeric substances (adhesives used by E. coli as a molecular glue). However, as the bacteria attempt to move away from the jagged nanopillar topography, shear forces are generated on the bacterial membrane. These forces cause separation of the inner membrane from the outer membrane, ultimately leading to bacterial death. Other experimental studies \cite{Hasan2013_6, Pogodin2013_10} have highlighted the role of bacterial cell membrane rigidity in determining susceptibility to mechanical rupture. Gram-negative bacteria, with their thinner membranes, are particularly sensitive to stretching induced by nanopillars. Differences in bacterial membrane architecture contribute to variations in bending rigidity. Gram-negative bacteria possess a cell envelope consisting of an outer and inner membrane separated by a thin (approximately 5 nm) peptidoglycan cell wall. In contrast, gram-positive bacteria have substantially thicker cell walls, ranging from approximately 20 to 100 nm \cite{Hazell2018_11, Silhavy2010_12}. This increased thickness imparts greater rigidity to gram-positive cell walls, reducing their susceptibility to stretching when exposed to nanopillared arrays. Beyond bacterial membrane properties, the configuration of nanopillared arrays—factors such as aspect ratio, density, and geometry—also significantly influences bactericidal efficiency \cite{Kelleher2016_3, Nowlin2014_9, Hazell2018_11, Hazell2018_14}. Furthermore, some experimental studies \cite{Rizeelo_2011_16, Rizzello2012_17} have reported that nanopillared surfaces can induce alterations in the genomic and proteomic profiles of bacteria. For example, when E. coli is incubated on single-walled carbon nanotubes, it exhibits elevated levels of stress response proteins associated with membrane damage and oxidative stress \cite{Kang2008_18}. An experimental and theoretical study by Linklater et al. \cite{Linklater2018_High} demonstrated that high-aspect-ratio nanopillars, such as carbon nanotube nanopillars, store energy due to bending. This stored energy is a significant factor in inducing physical rupture in both gram-positive and gram-negative bacteria. In another study, Linklater et al. \cite{Linklaater2022_Nanopillar} developed nanopillared polymer films for use as antibacterial packaging materials. These acrylic nanostructured films, featuring nanopillars with a height of 60 nm and a pitch/spacing of 30 nm, demonstrated superior antimicrobial and anti-biofouling behavior against gram-negative bacteria such as P. aeruginosa and gram-positive bacteria such as S. aureus. The films also exhibited a self-cleaning property, ejecting dead bacteria after killing them. Additionally, a recent review article by Linklater et al. \cite{Linklater2021_Mechano} explored the impact of nanoscale surface roughness on preventing bacterial colonization of synthetic materials. On flat, nanosmooth surfaces, bacterial adhesion involves an interplay of forces and physicochemical interactions, including van der Waals forces, electrical double-layer interactions, and electrostatics described by the extended Derjaguin–Landau–Verwey–Overbeek theory. In the presence of a nanopillared substrate, bacterial attachment is governed by the balance between adhesion energy and the deformation energy of the bacterial membrane. The ratio of membrane attraction to the nanopillar surface and membrane rigidity determines susceptibility to stretching-induced rupture. Despite these insights, conclusive evidence of bactericidal effects remains elusive, as no experimental study has successfully captured the bactericidal activity in real-time due to the small relevant length and time scales. Current experimental results are derived from post-contact analyses of bacteria or nanopillared substrates, leaving a gap in fully understanding the dynamic killing process.

All the aforementioned theoretical and experimental studies have attempted to solve the puzzle of bactericidal activity in the presence of nanopillared substrates incrementally, with each study contributing additional knowledge or bolstering previous hypotheses. However, we still observe varying degrees of predictability regarding the process of bacterial lysis. Experimental studies, with the current state of technology, remain incapable of capturing bactericidal activity due to the small spatial and temporal scales involved. Computational simulation techniques, such as Molecular Dynamics (MD), are well-known for their ability to investigate material behavior at these small scales, surpassing the limitations of experimental approaches \cite{Singh2022_19, Zhou2022_20, Singh2024_Templating, Singh2023_reliable}. For instance, Huang et al. \cite{Huang2008_ellshape} used MD simulations to create a quantitative physical model of the bacterial cell wall, predicting the mechanical response of cell shape to peptidoglycan damage and perturbation in the rod-shaped gram-negative bacterium E. coli. In another MD study, Salatto et al. \cite{Salatto2023_Structure} developed synergistic nanostructured surfaces made of polystyrene-block-polymethylmethacrylate with both bactericidal and bacteria-releasing properties. Salatto et al. conducted coarse-grained MD simulations of a flat lipid bilayer (a simplified model for E. coli) in contact with a substrate containing hexagonally packed hydrophilic nanopillars. They demonstrated that when bacterium-substrate interactions are strong, the lipid heads adsorb onto the nanopillar surfaces, causing the lipid bilayer to conform to the curvature of the nanopillars. This leads to high-stress concentrations within the membrane, particularly at the edges of the nanopillars, which act as a driving force for rupture. In finite element analysis (FEA) studies, Velic et al. \cite{Velic2021_Mechanics} suggested that nanopatterned surfaces do not primarily kill bacteria by rupturing them between protruding pillars, as previously thought. Instead, non-developable deformation near the pillar tips is more likely to create critical sites with significant in-plane strains, which can locally rupture and penetrate the cell. Velic et al.'s results also indicated that envelope deformation is increased by adhesion to nanopatterns with smaller pillar radii and spacing. The bacterial envelope was modeled as a mesh, including the outer membrane and cell wall, which are covalently linked by abundant lipoproteins to facilitate stress transfer. The inner membrane, being distant from the other layers, was considered non-load-bearing. Their study revealed that intrinsic adhesion forces could, in some cases (such as with "soft" envelope configurations), cause rupture of the bacterial envelope. Another FEA study by Cui et al. \cite{Cui2021_Validation} developed a three-dimensional, thin-walled bacterial cell model with turgor pressure. This model successfully simulated suspended bacterial cells and their interaction with flat and nanopillared surfaces. The results showed that the maximum stress and strain on bacterial cells adhered to the nanopillared substrate occurred at the liquid-cell-nanopillar three-phase contact line. Furthermore, it was found that the maximum stress and strain on a cell adhered to a flat surface were 2.7 MPa and 11.7\%, respectively, compared to 13 MPa and 65\% on a nanopillared surface, demonstrating the mechano-bactericidal properties of nanostructured surfaces. The computational studies conducted to date have provided further insights into the bactericidal activity of nanopillared substrates. However, a comprehensive understanding of the actual bactericidal mechanism remains elusive. Most computational studies have been unable to model dynamic bactericidal activity (in situ) using MD simulations, as the relatively large spatial scales (micrometer range) and long durations (microseconds) required to study these effects remain computationally challenging.

\begin{figure*}[hbt!]
\centering
\includegraphics[scale=0.75]{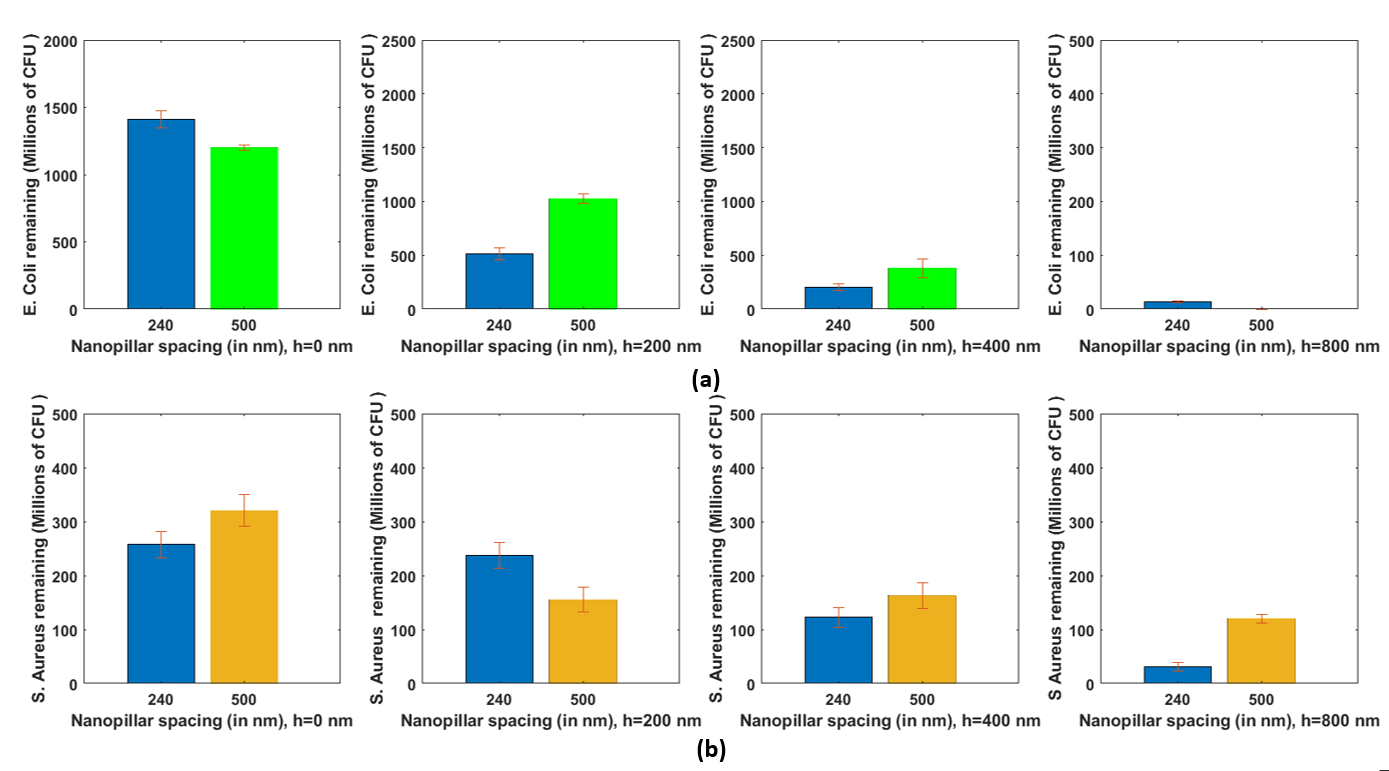}
\caption{Experimental results showing viable bacteria (gram-negative and gram-positive) with changes in nanopillar spacing and height: (a) For cylindrical bacteria E. coli, an increase in nanopillar height enhances the bactericidal effects in both cases—i.e., nanopillars with spacing of 240 nm and 500 nm. (b) For spherical bacteria S. aureus, an increase in nanopillar height also enhances the bactericidal effects for nanopillars with a spacing of 240 nm, although the increase is significantly less compared to cylindrical bacteria. Conversely, if the nanopillar spacing is increased to 500 nm, the bactericidal effects become negligible, and an increase in nanopillar height does not impact the bactericidal activity of the nanopillars. }
\label{fig:figureee1}
\end{figure*}

There is a lack of computational studies, particularly molecular dynamics (MD) simulations, that have developed a full-scale bacterial model with accurate membrane bending rigidity and analyzed the dynamic bactericidal activity of nanopillars. Therefore, this study aims to create a single-layer, coarse-grained bacterial membrane model that simulates the outer membrane, peptidoglycan layer, and inner membrane as a single layer, accurately capturing the bending rigidity of bacterial membranes in MD simulations. This simplified, full-scale bacterial model can simulate the dynamic bactericidal activity of nanopillared substrates while drastically reducing computational costs and accounting for the most critical mechanical property of bacterial membranes identified in previous studies—bending rigidity. Additionally, this study explores the potential influence of nanopillar height, spacing, and bacterial cell wall/membrane bending rigidity on bactericidal activity. Two coarse-grained models with distinct geometries have been developed to investigate these factors: First is a spherical model representing spherical bacteria, with a diameter of 500 nm, mirroring the geometry of S. aureus. Second is a cylindrical model measuring 1800 nm in length and 500 nm in diameter, representing gram-negative bacteria like E. coli. To understand the effect of variations in bacterial membrane bending rigidity, both models have been subjected to a range of cell wall bending rigidities from 15.1 $k_BT$ to 45.14 $k_BT$. The study also examines variations in nanopillar height (50–200 nm) and spacing/pitch (170–500 nm) to identify the most effective range of these parameters for bactericidal activity.

\section*{Experimental Testing}

Bio-implants, such as orthopedic implants, have a high risk of developing periprosthetic infections due to the growth of bacterial colonies on their surfaces. Both gram-positive and gram-negative bacteria, such as S. aureus and E. coli, can be the common culprits responsible for implant infections. Gram-positive bacterial cells are generally tougher and more resistant to mechanical damage due to their peptidoglycan cell walls being approximately 4 to 5 times thicker than those of gram-negative bacteria. In this study, we have developed polyimide nanopillared surfaces for bio-implants. For this, we devised a novel high-throughput fabrication technique for polyimide nanopillars (Appendix: Experimental Method) that combines top-down and bottom-up nanofabrication methodologies. This innovative approach enables the fabrication of large-scale arrays of polymer nanopillars with high aspect ratios and allows precise tuning of nanopillar geometries, facilitating the realization of optimal biomimetic designs. The fabrication process begins with the generation of a colloidal crystal mask across a wafer using monodispersed polystyrene nanospheres. Subsequently, an oxygen plasma etching step is employed to reduce the size of the nanospheres and establish uniform interstitial spacings. Next, a metal film is uniformly deposited over the surface, after which the nanospheres are removed, leaving behind a perforated mask. This mask is then utilized in a deep-silicon reactive-ion etching (RIE) process to create wells with high aspect ratios and vertical sidewalls. Afterward, the template is coated with polyamic acid, and vacuum annealing is performed to convert the oligomers into cross-linked polyimide. This facilitates the easy removal of the polyimide foil from the template, resulting in a standalone flexible substrate featuring high-density nanopillar arrays on its surface. Notably, this process allows for independent and precise control over nanopillar pitch, diameter, and height by adjusting the polystyrene diameter, oxygen RIE time, and the number of etching cycles in the deep-silicon RIE process, respectively. To assess the bactericidal properties of these polyimide nanopillared surfaces, we conducted experimental tests on E. coli and S. aureus to evaluate cell viability on the nanopillared surfaces. During these tests, the bacterial suspension was diluted in phosphate-buffered saline (PBS) to enable a quantitative analysis of how effectively these polymer films with various nanostructures could eliminate bacteria. More details on the experimental procedure can be found in the Methods section in the Appendix. We employed scanning electron microscopy (SEM) to examine the surface morphology of all samples before and after the bacterial cell viability tests. All the nanopillared surfaces developed for this study had a diameter of 50 nm.

Nanopillars with a high aspect ratio (length/diameter > 20) tend to break easily after testing, likely due to capillary effects during evaporation or gravitational collapse, thus limiting the maximum height of nanopillars created in this study. This observation also suggests the nonexistence of nanopillars with very large aspect ratios in nature, as they are mechanically less stable. For E. coli, a gram-negative bacterium, Fig. \ref{fig:figureee1}(a) illustrates the living population on different nanopillared surfaces with varying nanopillar spacing and heights after incubating them for 30 minutes. It can be observed that, in Fig. \ref{fig:figureee1}(a), the blue histogram (nanopillar spacing 240 nm) shows that as the nanopillar height increases from 0 nm (flat surface) to 800 nm, the remaining population of E. coli decreases drastically after 30 minutes of incubation. Specifically, for nanopillar heights of 200 nm, 400 nm, and 800 nm, the E. coli population is reduced by 63.71\%, 85.55\%, and 99.35\%, respectively. Similarly, in Fig. \ref{fig:figureee1}(a), the lime green histogram (nanopillar spacing 500 nm) indicates that as the nanopillar height increases from 0 nm (flat surface) to 800 nm, the remaining E. coli population decreases, though not as drastically as for the 240 nm spacing. Specifically, for nanopillar heights of 200 nm, 400 nm, and 800 nm, the E. coli population is reduced by 12.52\%, 68.37\%, and 99.91\%, respectively. These results suggest that, assuming constant nanopillar spacing and diameter, an increase in nanopillar height significantly enhances bactericidal efficiency. Conversely, increasing the nanopillar spacing, while keeping the same height and diameter, reduces bactericidal efficiency.

Similar studies were conducted on S. aureus, a gram-positive bacterium, after incubating it for 30 minutes on different nanopillared surfaces with varying spacing and height, as shown in Fig. \ref{fig:figureee1} (b). It can be observed in Fig. \ref{fig:figureee1} (b) that the blue histogram, representing a nanopillar spacing of 240 nm, shows a reduction in the remaining population of S. aureus as the nanopillar height increases from 0 nm (flat surface) to 800 nm. Specifically, the bactericidal efficiency for nanopillar heights of 200 nm, 400 nm, and 800 nm shows a reduction in the S. aureus population by 7.71\%, 52.34\%, and 87.72\%, respectively, after 30 minutes of incubation. Similarly, in Fig. \ref{fig:figureee1} (a), the yellowish-brown histogram, representing a nanopillar spacing of 500 nm, shows a reduction in the S. aureus population with increasing nanopillar height. However, this reduction is not as pronounced as that observed for the 240 nm spacing. For the 500 nm spacing, the reductions in the S. aureus population for nanopillar heights of 200 nm, 400 nm, and 800 nm are 51.56\%, 49.37\%, and 62.5\%, respectively. This suggests that, assuming the same nanopillar spacing and diameter, increasing the nanopillar height enhances bactericidal efficiency. Conversely, increasing the nanopillar spacing, assuming the same nanopillar height and diameter, decreases bactericidal efficiency. Additionally, it is evident that for the 500 nm spacing, the bactericidal efficiency remains nearly constant across the 200 nm, 400 nm, and 800 nm nanopillar heights. This implies that increasing the nanopillar spacing beyond a critical value, determined by the geometrical size of the bacteria, may diminish the effect of nanopillar height on bactericidal efficiency. Comparing the bactericidal efficiency of gram-negative and gram-positive bacteria, the results indicate that for the same nanopillar height, spacing, and diameter, the bactericidal efficiency is significantly higher for gram-negative bacteria. This observation suggests that the low bending rigidity of gram-negative bacterial membranes makes them more susceptible to damage on nanopillared surfaces. Furthermore, the study indicates that nanopillar height has a less pronounced effect on the bactericidal efficiency for the gram-positive bacterium S. aureus compared to the gram-negative bacterium E. coli, especially for larger nanopillar spacings (e.g., 500 nm spacing/pitch). The lower bactericidal activity observed in gram-positive bacteria with increasing nanopillar spacing can be attributed to the spherical shape and relatively small size of S. aureus cells, which measure less than 500-1000 nm in diameter. These cells are more likely to fit between the nanopillars without sustaining damage, unlike E. coli, which are larger. This highlights the critical role of bacterial geometry in the bactericidal activity of nanopillared surfaces. Given these observations, we hypothesize that bacterial survival is influenced not only by the robustness of their cell membranes but also by their size and geometry. The experimental results indicate that nanostructured surfaces can be designed to achieve optimal bactericidal efficiency while maintaining the mechanical stability of the nanopillars. However, the governing mechanisms behind the bactericidal activity of nanostructured surfaces remain unclear and are difficult to study solely through experimental testing, due to the time and length scales involved in bactericidal processes. Furthermore, identifying the best design for bactericidal nanostructured surfaces with combined optimal properties poses a challenge through experimental testing alone. Therefore, we decided to conduct computational simulations to explore the bactericidal mechanisms in greater depth and provide more insights into the interactions between bacteria and nanopillared substrates.

\section*{Computational Modeling}

\begin{figure}[h]
\centering
\includegraphics[scale=0.38]{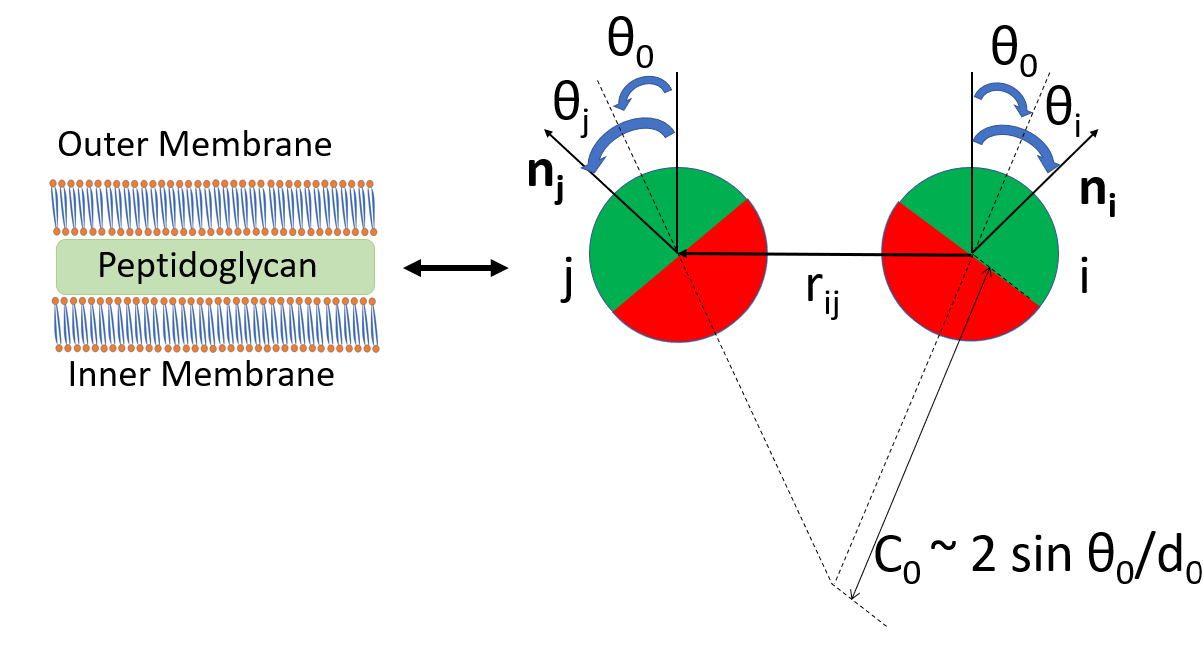}
\caption{The left figure illustrates the three different components of the bacterial membrane/cell wall: the outer membrane, the inner membrane, and the peptidoglycan layer located in the middle. The right figure depicts the coarse-grained (CG) model, where two beads collectively represent these three layers of the bacterial structure. Each CG particle is axisymmetric, with a particle-fixed unit vector $n$ representing the axis of symmetry and a mass of $m$. The schematic includes normal vectors $n_i$ and $n_j$ for CG particles $i$ and $j$, respectively. It also shows the angular parameters for the two beads, $(r_i, r_j)$, and their relationship with the angle $\theta_0$, which is a model parameter characterizing spontaneous curvature. Spontaneous curvature is denoted by $C_0$, while the average inter-particle distance between two coarse-grained atoms is represented by $d_0$. }
\label{fig:Image1}
\end{figure}

To understand and investigate the bactericidal effects of nanopillared surfaces, Molecular Dynamics (MD) simulations were employed in this study. All MD simulations were performed using LAMMPS \cite{LAMMPS}, a widely used MD simulation software. As shown in Fig. \ref{fig:Image1} (left), the bacterial membrane comprises three distinct components: the outer membrane, the inner membrane, and the peptidoglycan layer located in the middle. In this study, a one-particle-thick meshless model, i.e., a coarse-grained (CG) model, inspired by the studies of Yuan et al. \cite{Yuan2010_One} and Fu et al. \cite{FU2017_21}, was adopted to simulate the bacterial cell wall/membrane. The characteristic length of this model corresponds to the diameter of the CG particles, as shown in Fig. \ref{fig:Image1} (right), which matches the thickness of the lipid bilayer membrane. This approach enables the modeling of larger bacterial sizes in MD simulations, thereby reducing computational complexity and facilitating longer-duration simulations.

\noindent \textbf{Coarse-Grained Model for Bacteria's Membrane }

The utilization of a coarse-grained (CG) modeling approach for the bacterial bilayer membrane offers significant computational efficiency while preserving the membrane's essential physical property, specifically bending rigidity. This approach enables the exploration of characteristic length scales and simulation times that are considerably larger than those achievable with atomistic simulations. In this model, a single layer of coarse-grained particles, collectively representing the three layers of the bacterial membrane, was created. Since the thickness of the bacterial membrane is an order of magnitude smaller than the dimensions of the nanopillared structures, the bacterial membrane is assumed to be a thin elastic layer. Its structural details are neglected and are instead modeled using CG particles.

To simulate bacterial membrane behavior, a specialized lipid-lipid interatomic interaction potential was used, based on the previous work by Yuan et al. \cite{Yuan2010_One}. This unique inter-particle potential encompasses three distinct interactions designed for CG lipid mesoscopic particles: head-head, tail-tail, and tail-head interactions. These interactions account for both bonding and angular interactions between pairs of CG atoms within the bacterial membrane. This interparticle potential is constituted of two functions, radial interparticle potential function  $U(r) = u_R(r) + u_A(r)$ and angular interparticle potential function $\phi(\hat{r}_{ij},n_i,n_j)$ which describes the distance and orientation dependencies of CG particles respectively. The schematic in Fig. \ref{fig:Image1} (right) represents a generic relative position and orientation of such a particle pair, where the two halves of each particle are colored distinctly to indicate its orientation. The center position vectors of particles $i$ and $j$ are represented as $r_i$ and $r_j$ respectively. The interparticle distance vector is then $r_{ij} = r_i - r_j $. Also, we define the unit vectors $n_i$ and $n_j$ to represent the axes of symmetry of particles $i$ and $j$, respectively. For simplification, the rotational degree of freedom about the axis of symmetry of each particle is neglected. Thus, each coarse-grained particle has five degrees of freedom, i.e. two rotational and three translational. Fig. \ref{fig:Image1} (right) also represents the concept of spontaneous curvature ($C_0$) between two CG particles. Here, $d_0$ represents the mean interparticle distance between two coarse-grained atoms, $\widetilde{C_0} = R_0C_0$ is the dimensionless instantaneous curvature, and $R_0$ corresponds to the radius of the spherical CG atom. The orientations of atom pairs $(r_i, r_j)$ are denoted as $(\theta_i, \theta_j)$, and the position of the $i^{th}$ coarse-grained atom is indicated by $r_i$. Radial inter-particle potential is utilized to describe the effects of covalent bonds and can be further subdivided into repulsive and attractive parts, designated as $u_R(r)$ and $u_A(r)$, respectively. These components are defined as follows.

\begin{equation}  
           u_R(r)=\epsilon \{(\frac{r_{min}}{r})^4-2(\frac{r_{min}}{r})^2\}
 \end{equation}  
 \begin{equation}  
           u_A(r)=\epsilon\cos^{2\zeta}\{\frac{\pi}{2}(\frac{r-r_{min}}{r_c-r_{min}})\}
 \end{equation}  
 
 \noindent where $u_R(r)$ represents the repulsive branch of radial interparticle potential and adopts the 4-2 type Lennard Jones (LJ) potential function. 4-2 LJ potential is adopted as it has a lower restoring force than 12-6 LJ potential at equilibrium distance.  $\epsilon$ and $\sigma$  are LJ potential function parameters \cite{Wang_2020_23} respectively.  $\epsilon$ for LJ potential represents the depth of the potential well also known as dispersion energy. $\sigma$ is the distance at which the particle-particle potential energy is zero and simply put it represents the size of the particle. $r=|r_{ij}|= |r_i-r_j|$ is the instantaneous distance between atom $i$ and atom $j$. $u_A(r)$ represents the attractive branch of radial interparticle potential. $u_A(r)$ is a cosine function that smoothly decays to zero at $r_{c}$. 
$r_{c}$ is the cutoff radius of the radial interparticle potential. $\zeta$ tunes the shape of the attractive branch of the interparticle potential function hence the diffusivity of the particles,  $r_{min}$ is the distance that minimizes the potential energy function $u_A(r)$, The attractive branch and repulsive branch meet at $r = r_{min}$ with $C^1$ continuity where $r_{min}= 2^{1/6}\sigma$, the same as in 12-6 LJ potential.
 
 The angular interparticle potential function represents the relative orientation between two CG atom pairs ${r_i, r_j}$ and is defined as follows: 
 
 \begin{equation}  
           \phi(\hat{r}_{ij},n_i,n_j)=1+\mu(a(\hat{r}_{ij},n_i,n_j)-1) 
 \end{equation}  
  \begin{equation}  
           a(\hat{r}_{ij},n_i,n_j)= (n_i \times \hat{r}_{ij}).(n_j \times \hat{r}_{ij}) + \sin{\theta_o}(n_i-n_j).\hat{r}_{ij}-\sin^2{\theta_o}
 \end{equation}

\noindent where $\hat{r}_{ij}={r}_{ij}/r$ is the unit position vector between atoms $i$ and $j$, $\phi$ is used to weigh the interaction strength in different relative orientations, leading to the final form of the anisotropic pair potential. This follows from the anisotropic potentials for liquid crystal or colloids from the studies of Gay et.al. and Everaers et.al.  \cite{Gay1980_Modification, Everaers2003_interaction}. $\theta_o$ parameter is the instantaneous curvature between two coarse-grained atoms. The parameter $\mu$ weighs the energy penalty when the particles are disoriented from $\theta_o$ and is therefore related to the bending rigidity of the model membrane. The total inter-particle potential energy function $U$ can be expressed in terms of $\phi, u_R(r) $ and $u_A(r)$ and is defined as follows:

  \begin{equation}  
           U({r}_{ij},n_i,n_j)= 
           \begin{cases}
     u_R(r)+ [1-\phi(\hat{r}_{ij},n_i,n_j)]\epsilon, &  r< r_{min}\\
     u_A(r)\phi(\hat{r}_{ij},n_i,n_j), & r_{min}<r<r_c
      
    \end{cases} 
 \end{equation}

When applying the previously mentioned equations, the simplest scenario arises when the normal vectors $n_i$ and $n_j$ are aligned in parallel. In this case, the values of $a$ and $\phi$ become 1. This represents a planar membrane. This potential function in LAMMPS \cite{LAMMPS} requires six parameters that must be specified in the input script for LAMMPS in the following order: $\epsilon$, $\sigma$, $r_{cut}$, $\zeta$, $\mu$, and $\sin \theta_o$ to accurately define the parameters of bacteria's membrane.\\

\noindent\textbf{Coarse-Grained Model for Outer Fluid and Cytoplasm } 

There are two types of fluid molecules in our model: the internal liquid within the bacterial lipid cell membrane, corresponding to the cytoplasm of the bacteria; the external water molecules surrounding the bacterial cell. For modeling the outer fluid and cytoplasm, Lennard Jones (LJ) inter-particle potentials have been employed in this study. These potentials help describe the interactions between various components, including lipid membrane-water, cytoplasm - cytoplasm, cytoplasm - membrane, cytoplasm - nanopillar substrate, and membrane-nanopillar substrate. The 12-6 LJ potential is mathematically defined by the following equation:

  \begin{equation}  
           E_{LJ}= 4\epsilon[(\frac{\sigma_{eq}}{r_{ij}})^{12}-(\frac{\sigma_{eq}}{r_{ij}})^{6}]
 \end{equation}

\noindent where $r_{ij}$ represents the separation between two coarse-grained water atoms, $\sigma$ denotes the equilibrium length in the 12-6 Lennard-Jones interactions, and $\epsilon$ characterizes the depth of the potential energy function, which relates to the strength of interaction between the particles. To determine the mass of each coarse-grained membrane and cytoplasm atom, we used the approximate density of either gram-positive bacteria, such as S. aureus, or gram-negative bacteria, like E. coli and volume occupied to evaluate the mass of individual coarse grained atom for membrane and cytoplasm respectively. The mass of each water molecule was derived based on the assumed density of water, approximately 1 gram/cc. Additionally, the mass of the nanopillared array was determined by assuming the composition of polyimide on the basis of the studies of Yi et.al. \cite{Yi2023_Asmart}. It's worth noting that the interaction between the nanopillared substrate and the membrane was considered attractive (since polyimide nanopillars are hydrophobic where $\Delta G_{iwi}$ is negative). $\Delta G_{iwi}$ is a parameter that explains hydrophobicity as the free energy of interaction between two materials \cite{VanOss_etal,Elfazazi2021_etal}).  To mimic the behavior of real polyimide nanopillars and E. coli bacteria (strain MG1655), Yi et.al \cite{Yi2023_Asmart} has evaluated the contact angle of water on 200 nm long nanopillars and found it to be approximately equal to 67.8 $\pm$ 2.19 degrees (suggesting that polyimide nanopillars are hydrophobic). Following the studies of Elfazazi et.al. \cite{Elfazazi2021_etal} $\Delta G_{iwi}$ can be approximated to be equal to -24.34 $mJm^{-2}$ for the interaction between E. coli bacteria (strain MG1655) with polyimide nanopillars. Current LJ interatomic interaction used for the interaction between nanopillar and bacteria $\Delta G_{iwi}$ =-0.0093 $mJm^{-2}$ is low negative $\Delta G_{iwi}$ which indicates a very small attractive interaction between nanopillars and bacterial membrane. Although, this LJ interaction represents a very low value compared to $\Delta G_{iwi}$ = -0.00963 $mJm^{-2}$ estimated from experimental study. To compensate for this we use an additional force on bacteria i.e loading rate to simulate the attractive forces not accounted for using the LJ interactions. We wanted to emphasize that we have deliberately chosen a very low value of attractive interaction between polyimide nanopillars and bacterial membrane as the bacteria - nanopillar interaction constitutes of multiple components including van der waals interaction, electrostatic interactions, surface energy interactions etc., therefore approximating the total attractive interactions with LJ interatomic potentials might be unnatural. Low attractive interaction will ensure the bacterial killing mechanism is mechanically induced via a loading rate or force on bacteria. This will allow the bacteria to slide/navigate  on nanopillared surfaces and avoid getting killed instantaneously, which is what we observe from most experimental studies. To the best of our knowledge till date  no study has simulated a whole bacterium dynamic activity  with nanopillar interaction/adhesion for polyimide material using E. coli.  \\

\noindent\textbf{Coarse-grained Molecular Dynamics Simulations } 

\begin{figure}[ht]
\centering
\includegraphics[scale=0.6]{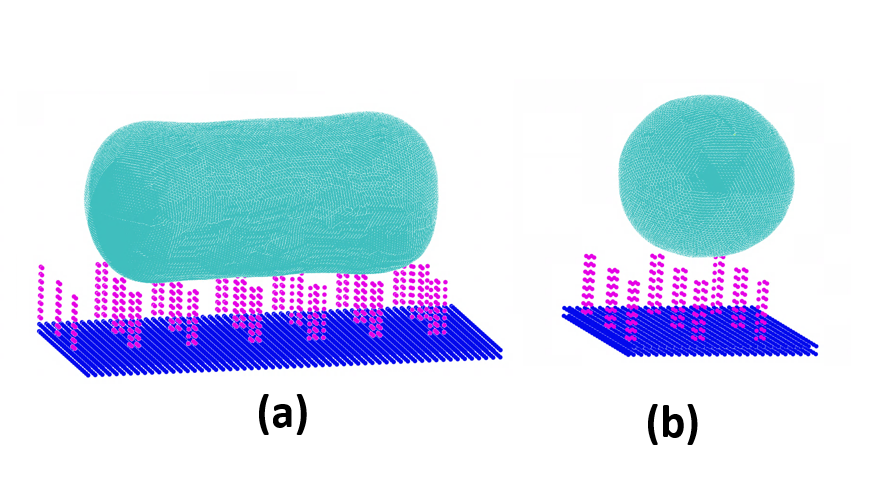}
\caption{Coarse grained MD simulation model (a) Cylindrical E. coli with diameter 500 nm and tip to tip length 1800 nm (b) Spherical S. aureus with diameter 500 nm}
\label{fig:figure3}
\end{figure}

As shown in Fig. \ref{fig:figure3}, two different CG models have been developed to simulate bacteria with different geometry. First is a cylindrical shaped bacteria mimicking E. Coli  that is 1800 nm long and has a diameter of 500 nm as shown in Fig. \ref{fig:figure3} (a), and second is a spherical shaped bacteria mimicking S. aureus that is 500 nm diameter as shown in Fig. \ref{fig:figure3} (b).

In MD simulations, the initial configuration of the CG models were set up in a way that bacterial cell, suspended in the surrounding water molecules, is separated from the nanopillared 
 surface by at least 100 \AA. To remove any excessive internal stresses in the cell membrane, the effective radius of the coarse-grained water molecules was adjusted in the initial configuration by modifying the equilibrium bond length and ensuring zero initial harmonic bond energy in the cell membrane. To accomplish this, a modified harmonic bond function, as developed by Fu et al. \cite{FU2017_21}, was employed in LAMMPS. The modified harmonic bond function utilized the initial configuration denoted as $x_o$ to calculate the bond length $l_o(r_i, r_j)$ between each pair of atoms at the beginning of the simulation. This approach differed from the built-in harmonic function present in LAMMPS, where the bond length $r_o$ remained constant. The modified harmonic bond energy function can be expressed as follows:

  \begin{equation}  
           E_{har}= K(r_{ij}-l_o(r_i,r_j))^2
 \end{equation}  
 \noindent where $l_o $ is the initial bond length.

%\begin{table}[hbt!]
%\caption{\label{tab:tab2} Lennard Jones parameters used in MD simulations  }
%\centering
%\begin{tabular}{lccc}
%\hline
% 0.02 \sigma_{m-w} & 240 & \epsilon_{m-w}  \\
% 0.2 \sigma_{m-p} & 240 & \epsilon_{m-p}  \\
% 0.0002 \sigma_{w-p} & 240 & \epsilon_{w-p}  \\
%\hline
%\end{tabular}
%\end{table}

Our CG spherical membrane model consisted of 8,348 CG membrane atoms, replicating a 500 nm diameter membrane similar to that of S. aureus, as illustrated in Figure \ref{fig:figure3} (b). The entire simulation box encompassed a total of 29,094 atoms, including the cytoplasm, outer water, the membrane, and the nanopillared substrate. The simulation box used in this study has a length, width and height of 6940 x 9140 x 7250 \AA$^3$. Similarly, our CG cylindrical membrane model consisted of 21,468 CG membrane atoms, replicating a 500 nm diameter and 1800 nm long cylindrical membrane resembling that of E. coli, as illustrated in Figure \ref{fig:figure3} (a). The entire simulation box encompassed a total of 67,937 atoms, including the cytoplasm, outer water, membrane, and the nanopillared substrate. The simulation box used in this study has a length, width, and height of 6940 x 9140 x 16160 \AA$^3$. We used a fluid membrane pair potential \cite{FU2017_21} to define interatomic interactions between CG atoms that make up the bacterial cell membrane. For all other interatomic interactions, we utilized the Lennard Jones pair potential. The Fluidmembrane pair potential relied on six different parameters: $\epsilon$, $\sigma$, $r_{cut}$, $\zeta$, $\mu$, and $\sin\theta_o$ (e.g., for bacteria with low membrane bending rigidity, the parameters were set to 0.2, 103, 200, 4, 5, and 0). All simulations were conducted with a time step of 0.1 femtoseconds. The modeling process began with energy minimization to alleviate internal stresses, employing a conjugate gradient algorithm in LAMMPS. Subsequently, MD simulations were performed in two substeps. In the first substep, we used the NPT ensemble to equilibrate the system at 300 K for 200 ps before transitioning to the NVT ensemble. In the second substep, an additional acceleration of 0.0002 eV/(m x \AA) per CG atom, also known as loading rate ( L.R.), was applied to the bacterial membrane and internal cytoplasm to simulate the effects of gravity, attractive forces between bacteria and nanopillars not accounted for in the nanopillar bacterial membrane interaction and hydrodynamic forces, where m is the mass of CG atom. For this study, we designed the nanopillar array to mimic the configuration found in cicada wings, featuring a nanopillar diameter of 60 nm, a nanopillar spacing of 170 nm between two nanopillars, and a nanopillar height of 200 nm. Throughout our investigation, we kept the nanopillar diameter constant, while we varied the nanopillar height, spacing and bending rigidity of bacterial membrane as our parameters of interest, as shown in Table \ref{tab:tab3}. The terms used to describe the nanopillar configuration include spacing (the distance between two nanopillars), height (the distance from the substrate surface to the tip of the nanopillar), and diameter (the width of the cylindrical nanopillar).

\noindent\textbf{Model Calibration}

The bending rigidity of the bacterial cell membrane primarily governs the mechanical response of bacteria on nanostructured surfaces. It is critical for the developed CG model to capture the correct bacterial membrane bending rigidity  for understanding its failure mechanism on nanostructured surfaces. Therefore, the developed CG model is calibrated to match the bending rigidity of simulated bacteria in this study. In the absence of surface tension, the bending rigidity can be determined by examining the power spectrum of height fluctuations of a free-standing planar membrane, also referred to as undulatory motions, denoted as $<|h(q)|^2>$. This is defined as follows, according to the research by Fowler et al. \cite{Fowler2016_14}:

  \begin{equation}  
           <|h(q)|^2>= \frac{K_B T }{K_c q^4}
 \end{equation}

\begin{figure}[ht]
\centering
\includegraphics[scale=0.20]{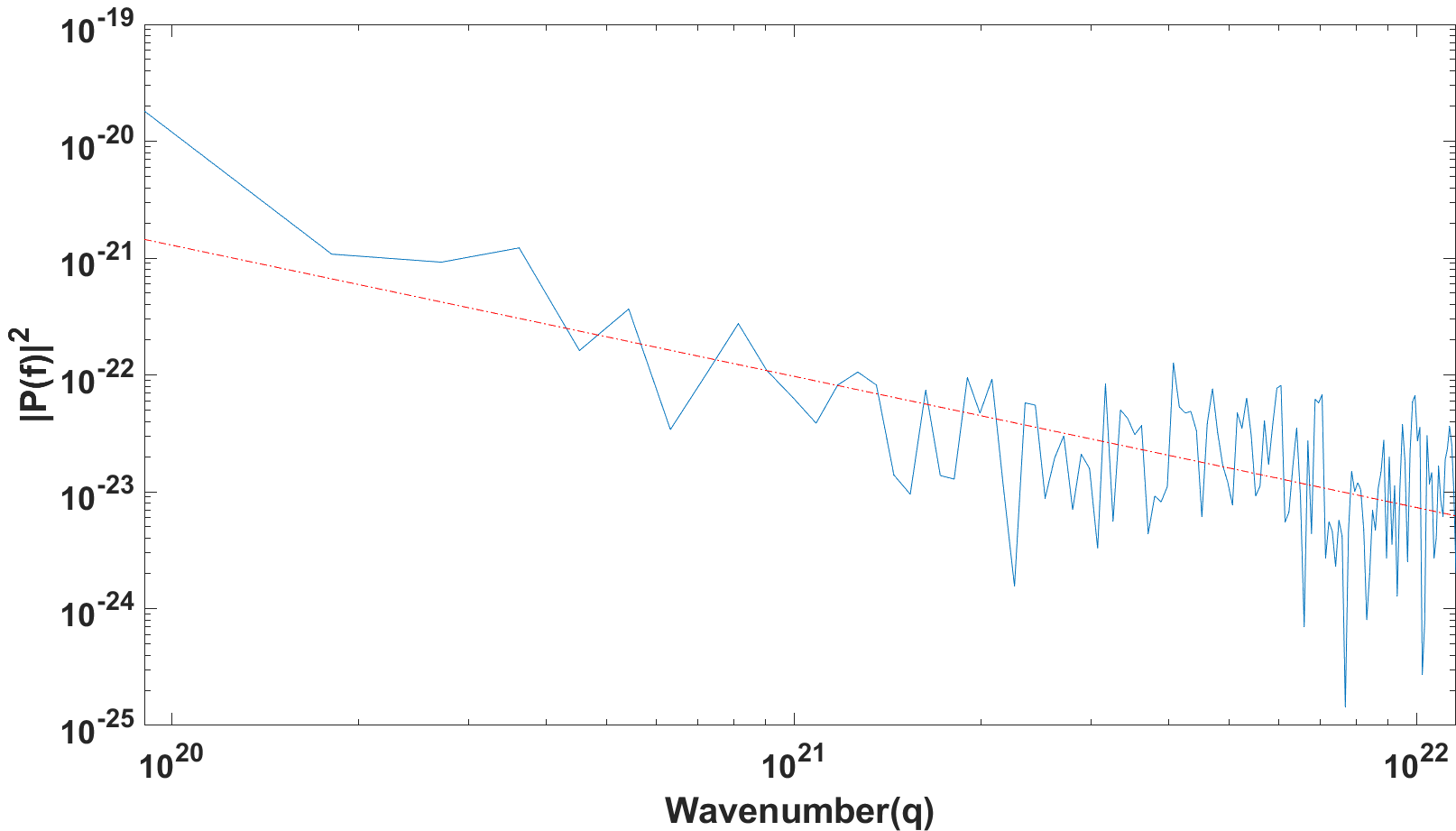}
\caption{Power spectrum of height fluctuations of planar membrane vs wavenumber. Red dashed line is an approximate linear fit to evaluate bending rigidity}
\label{fig:figure2}
\end{figure}

\noindent where q is the magnitude of  wavevector (i.e wavenumber in $nm^{-1})$), $K_c$ is the bending rigidity, T is the temperature in Kelvin and $K_B$ is the boltzmann's constant. This equation was firstly introduced by Helfrich  \cite{Helfrich1973_15} and is only applicable for lipid bi-layers where length scales are much longer than  thickness of bi-layer membrane. The power spectrum of height fluctuation of atoms in a planar lipid bi-layer membrane can be calculated in MD simulations, which can be fitted into the above equation to evaluate bending rigidity of the planar membrane a shown in Fig. \ref{fig:figure2}

To calibrate the developed CG membrane model, we employed a planar membrane model measuring 3000 \AA  \space  x  3000 \AA, consisting of 2500 particles. Bacterial membrane bending rigidity  parameters  $\mu$ and $\epsilon$ as shown in Eq. 5 were tuned to get the accurate bacterial membrane bending rigidity. This planar membrane simulation featured periodic boundary conditions in in-plane direction of the membrane and non-periodic boundary conditions in out of plane direction, effectively mimicking an infinitely long membrane. This planar membrane model allowed us to adopt the Helfrich model for estimating the bending rigidity of the membrane as in this planar membrane model the characteristic length of the planar membrane is more than ten times of $\sigma$, where $\sigma$ represents the thickness of membrane. Then energy minimization was conducted to planar membrane to get rid of any residual stresses. After initial energy minimization, the planar membrane was subjected to the Number of atoms, Volume and Temperature constant (NVT) ensemble at 300 K for 100 ps to achieve equilibrium. Following this, the simulation switched to the Number of atoms, Volume and Energy (NVE) constant ensemble for an additional 100 ps. The height fluctuations spectra of the simulated planar membrane is analyzed using the Fast Fourier Transform technique in MATLAB \cite{MATLAB2010_16} during NVE ensemble. Subsequently, we plotted the fluctuation spectra against $q^{-4}$ to determine the bending rigidity of the membrane. As mentioned above, $\mu$ and $\epsilon$ parameters in Eq. 5 were tuned using trial and error method to match the bending rigidity of bacterial membrane.  Previous research has indicated that gram-negative bacteria, like E. coli, possess weaker cell walls with a bending rigidity of the order of 13 $\pm$ 5 $K_bT$. In contrast, gram-positive bacteria such as S. aureus have bacterial cell membranes with higher bending rigidity i.e. 43 $\pm$ 5 $K_bT$. The parameters in our CG membrane models were calibrated to match the bending rigidity of gram-negative bacteria such as E. Coli which has an approximate membrane bending rigidity of 13 $\pm$ 5 $K_bT$, and similarly gram-positive bacteria such as S. aureus, which has an approximate membrane bending rigidity of 43 $\pm$ 5 $K_bT$.

\begin{figure*}[hbt!]
\centering
\includegraphics[scale=0.70]{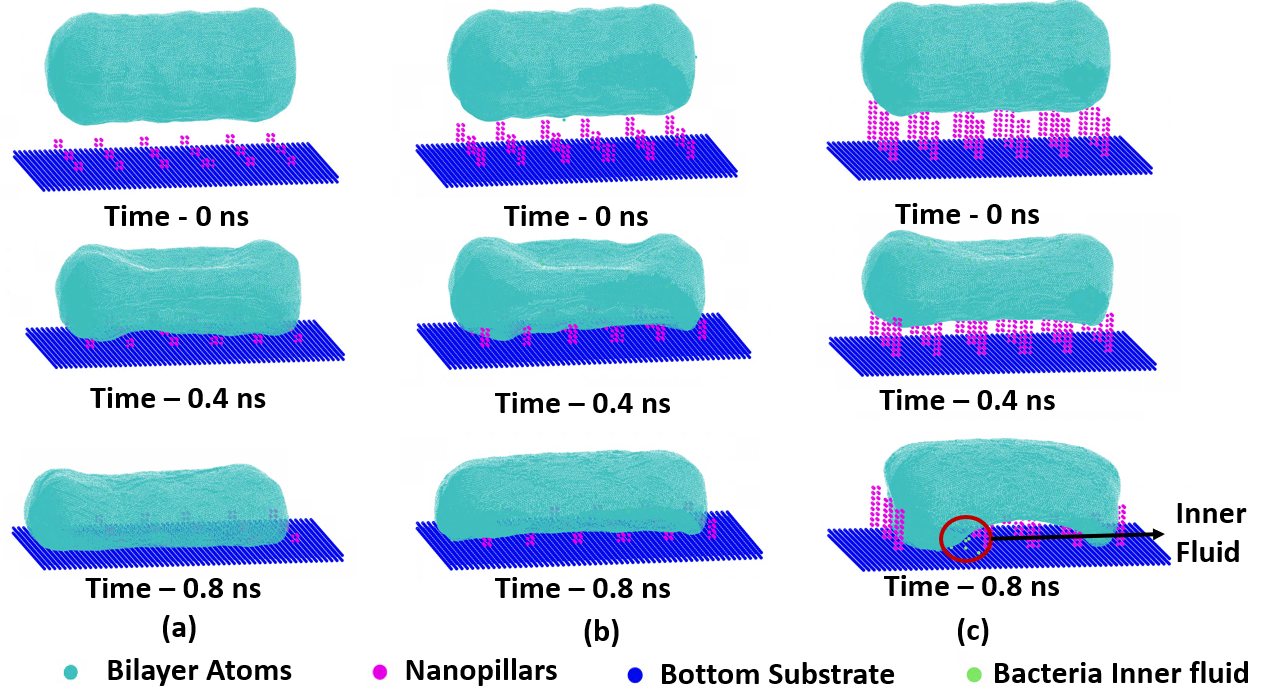}
\caption{Simulation of bacteria cell with low bending rigidity membrane  and varying nanopillar height (a) with nanopillar height = 50 nm (b) with nanopillar height = 100 nm (c) with nanopillar height = 200 nm. It can be noticed that with the progression of time for cases with bacteria and nanopillared surfaces with height 50 nm and 100 nm, bacteria wrapped around the nanopillars and no damage to the bacterial membrane can be observed after 0.8 ns. But for case with  bacteria and nanopillared surface with height 200 nm, bacteria membrane tried wrapping around but eventually gets teared down as the bacteria membrane could not reach the bottom surface. }
\label{fig:figuree5}
\end{figure*}

\begin{figure*}[hbt!]
\centering
\includegraphics[scale=0.70]{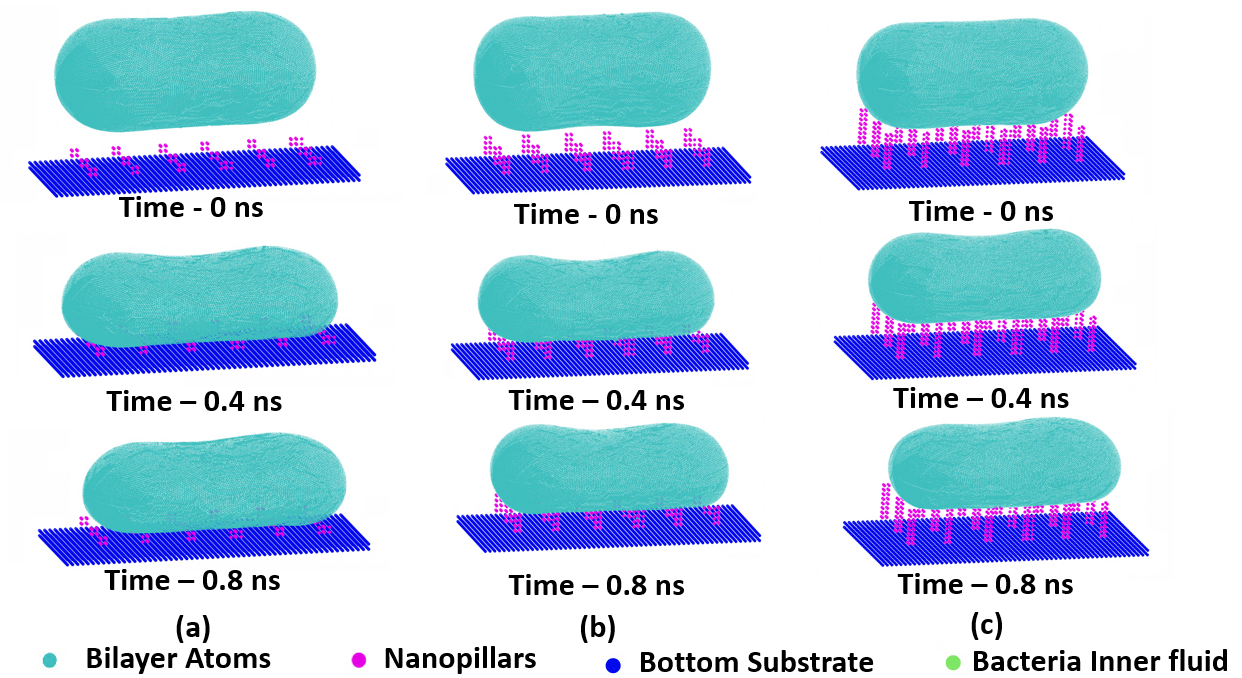}
\caption{Simulation of bacteria cell with high bending rigidity membrane and varying nanopillar height (a) with nanopillar height = 50 nm (b) with nanopillar height = 100 nm (c) with nanopillar height = 200 nm. It can be noticed that with the progression of time for all the cases  bacteria sat on top of nanopillars and no damage to the bacterial membrane can be observed after 0.8 ns}
\label{fig:figuree6}
\end{figure*}

\section*{Results and Discussions}
This section discusses CG MD simulation results for investigating the bacteria-nanopillared substrate interaction to understand the bactericidal mechanisms of nanostructured surface. This study examines different bacterial varieties in terms of shapes, i.e. rod vs spherical, as well as cell wall thickness, i.e. gram-negative vs gram-positive. We also investigate how the arrangement of nanopillars including the radius, height and spacing impact the bacteria-nanopillar interaction. We compare the CG MD simulation results with the experimental data for model validation.

\begin{figure*}[hbt!]
\centering
\includegraphics[scale=0.72]{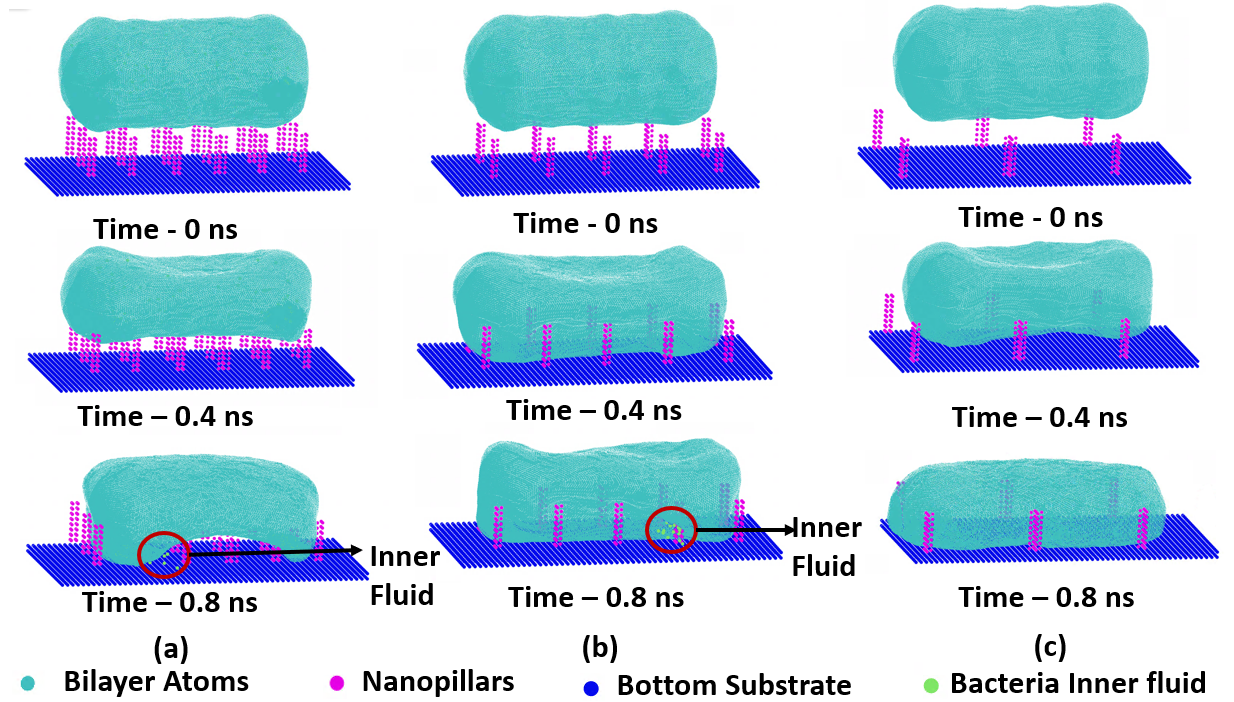}
\caption{Simulation of bacteria cell with low bending rigidity membrane and varying nanopillar spacing (a) with nanopillar spacing = 170 nm (b) with nanopillar spacing = 240 nm (c) with nanopillar spacing = 500 nm. It can be noticed that for highest spaced nanopillars bacteria squeezed in between the array and survived but for other cases the nanopillar membrane was stretched and torn out. }
\label{fig:figuree7}
\end{figure*}

\begin{figure*}[hbt!]
\centering
\includegraphics[scale=0.7]{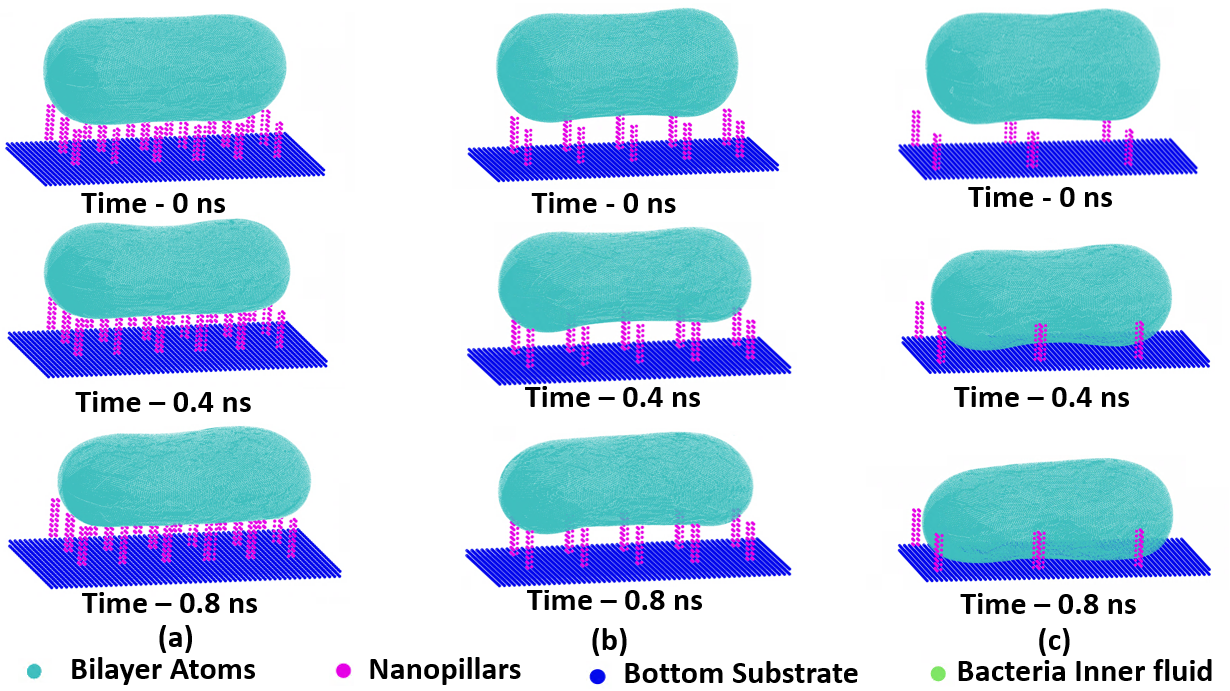}
\caption{Simulation of bacteria cell with high bending rigidity membrane and varying nanopillar spacing (a) with nanopillar spacing = 170 nm (b) with nanopillar spacing = 240 nm (c) with nanopillar spacing = 500 nm. It can be noticed that for all the cases bacteria was able to survive i.e. bacteria for low nanopillar spacing the bacteria was able to sit on top of nanopillars and for high membrane bending rigidity the bacteria was able to sink to the surface and survive}
\label{fig:figuree8}
\end{figure*}

Bacteria can be classified into five groups by their basic shapes: rod, spherical, spiral and comma. In this study, we look into two of them, i.e. rod and spherical. In addition, bacteria have different cell-wall structures that put them into two basic categories: gram-negative and gram-positive. Gram-positive bacteria have a thicker cell wall compared with gram-negative bacteria. In our experimental study, the nanostructured surface shows much higher efficacy in killing gram-negative bacteria compared with gram-positive bacteria. Therefore, in our CG MD simulations, we custom the bending rigidity of the bacterium model to simulate gram-negative bacteria (15 $K_BT$) and gram-postive bacteria (45 $K_BT$).

\subsection*{Rod-shaped Bacteria}
Fig. \ref{fig:figuree5} show the responses of rod-shaped bacteria with a bending rigidity 15 $K_BT$ on the nanostructured surfaces with aligned nanopillars spacing at 170 nm. The bending rigidity 15 $K_BT$ is chosen based on literature results to mimic the mechanical property of cell wall membrane of gram-negative bacteria \cite{Wong2019-km} like E. Coli. The height of nanopillars are varied from 50 nm, to 100 nm, and 200 nm while the bacterium is 1 micrometer long with a radius of 500 nm , as shown in Fig. \ref{fig:figuree5} (a), (b), and (c). 

CG MD simulations starts with the bacterium floating in water at time 0. The water molecules are removed for a better visualization. It can be observed that as the bacterium approached the substrate, it flatted out a little bit due to the adhesive interaction between the bacterium and the nanopillars. After the bacterium came into contact with the nanopillars, the cell wall suspended between nanopillars and the cell began to deform gradually. With the nanopillars of 50 nm tall in Fig. \ref{fig:figuree5} (a), the bacterium can be found to fully sink down to the substrate after 0.8 ns. The cell membrane sagged between the nanopillars and eventually wrapped around the nanopillars along the edge of the bacterium. The bacterium had the cell wall touch the bottom of the substrate around the edge of the bacterium, which stopped a further deformation of the bacterium. As the height of nanopillars increased to 100 nm, the bacterium has the cell wall suspended and sagged between nanopillars with the largest sagging happened at the two ends of the bacterium. It can be noted that the cell wall of the bacterium kept sagging into the space between nanopillars until it touched the bottom which can create tension in the cell membrane. By further increasing the height of the nanopillar to 200 nm as in Fig. \ref{fig:figuree5} (c), the bacterial membrane sagged and suspended between the nanopillars in the middle of bacterium while the two ends can be observed to sink down to the substrate after 0.8 ns. As the two ends of the rod bacterium dropped to reach the bottom of the substrate due to adhesion, it leads to an excessive stretching in the cell membrane, which is believed to eventually tear the bacterial membrane. It can be observed that increasing the height of nanopillars can lead to amplified membrane deformation while the bacteria sinked down to the bottom of the substrate due to adhesion. The low bending rigidity of gram-negative bacteria allows the membrane to sag between nanopillars induced by the adhesive interaction, which can result in tension and stretching within the membrane. Nanostructured surface like nanopillars can increases the contact surface between the bacteria and the substrate. The increase of nanopillar height can quickly increase the contact surface. With a spacing of 170 nm, it can be found that the bacterium can sag and stretch the cell membrane on 50 nm tall of the nanopillars to touch the bottom without tearing the membrane. As the height increased to 200 nm, the significant sagging of cell membrane lead to extensive tension and finally rupture the membrane.   

%These finding indicates that nanopillar induced deformation is critical for killing the bacteria and also there is a critical nanopillar height threshold (nanopillar height should be more than bacterial height in horizontal condition), approximately around 200 nm for this particular study, to induce enough deformation in cell membrane of gram-negative cylindrical bacteria with low membrane bending rigidity (15 $K_BT$). Notably, this aligns with experimental outcomes presented in Fig. \ref{fig:figureee1}, which showcased a decrease in the population of viable E. coli with increasing nanopillar height. This mechanism echoes with previous research findings, including those by Ivanova et al. and Pagodin et al. \cite{Ivanova2013_2,POGODIN2013_17}, who emphasized the failure mechanism in which bacterial membrane sags between nanopillars and tears apart. 

\begin{figure*}
\centering
\includegraphics[scale=0.7]{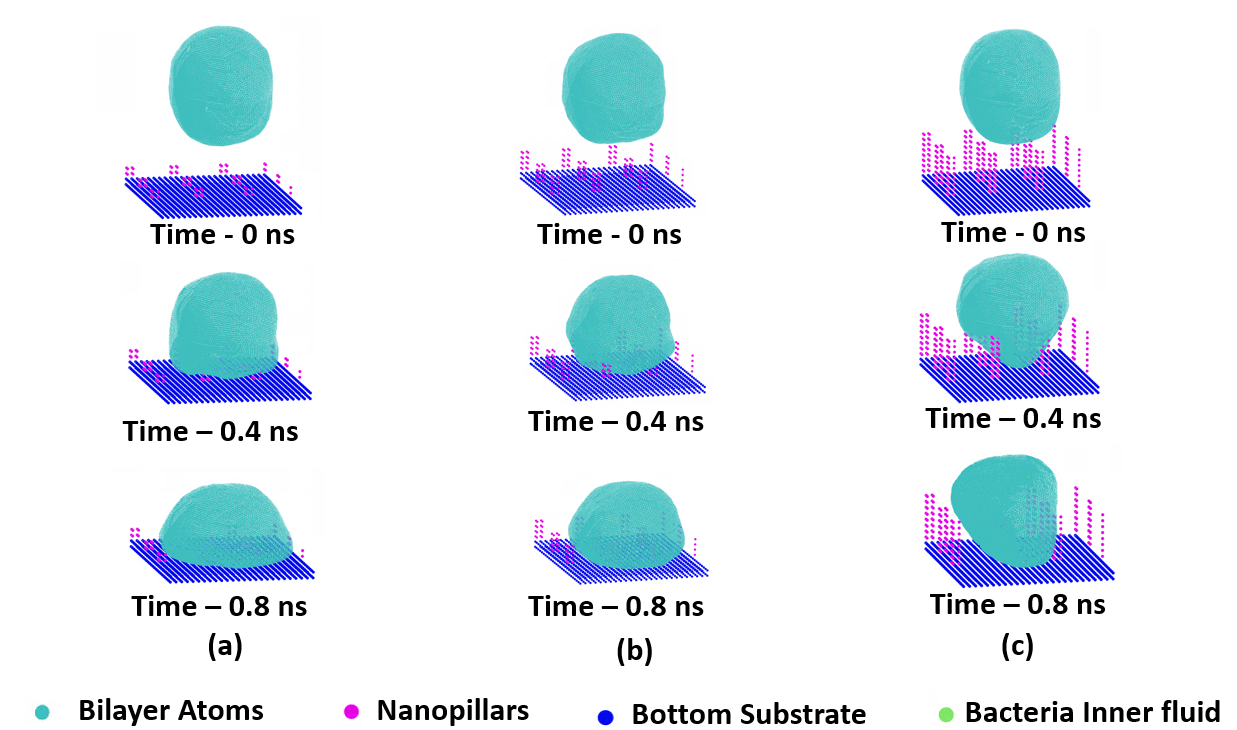}
\caption{Simulation of bacteria cell with low bending rigidity membrane  and varying nanopillar height (a) with nanopillar height = 50 nm (b) with nanopillar height = 100 nm (c) with nanopillar height = 200 nm. It can be noticed that with the progression of time  nanopillared surfaces with height 50 nm, 100 nm and 200 nm, bacteria wrapped around the nanopillars and no damage to the bacterial membrane can be observed after 0.8 ns. But it can also be observed that with increasing nanopillar height the deformation in bacteria membrane is increasing}
\label{fig:figuree9}
\end{figure*}

\begin{figure*}
\centering
\includegraphics[scale=0.72]{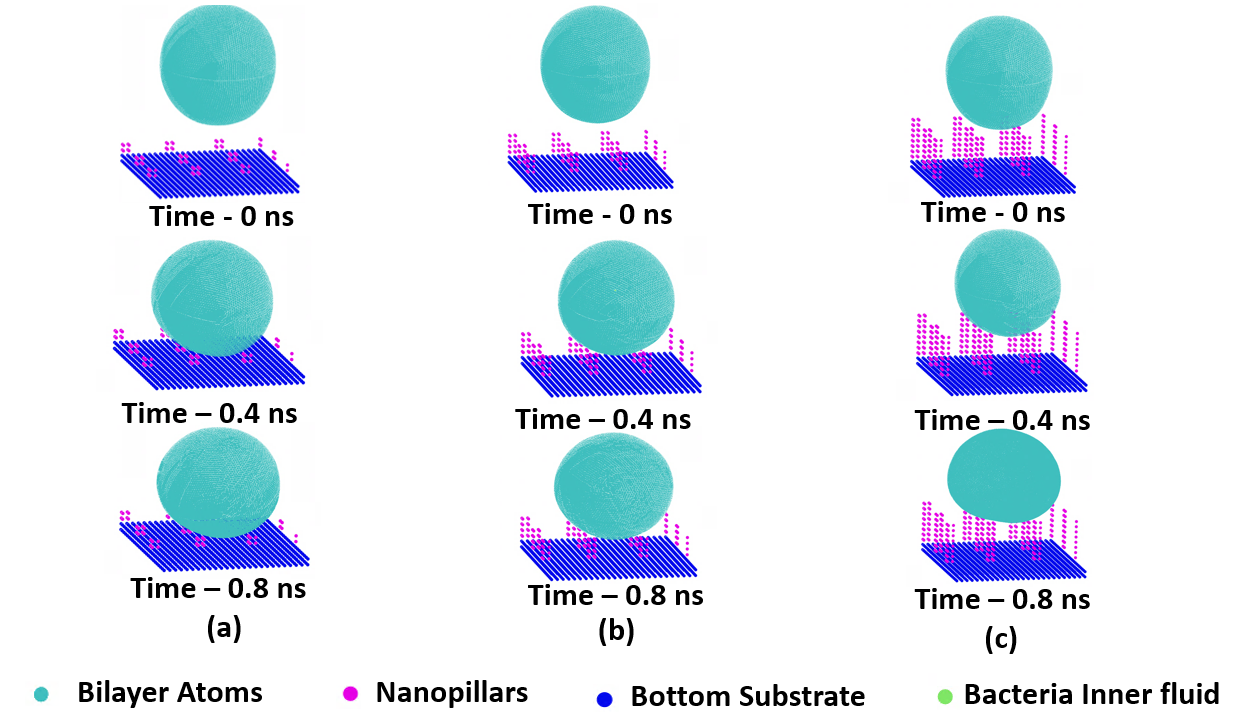}
\caption{Simulation of bacteria cell with high bending rigidity membrane and varying nanopillar height (a) with nanopillar height = 50 nm (b) with nanopillar height = 100 nm (c) with nanopillar height = 200 nm. It can be noticed that with the progression of time and for all the cases spherical bacteria sat on top of nanopillars and no damage to the bacterial membrane can be observed after 0.8 ns}
\label{fig:figuree10}
\end{figure*}

Bacteria with bending rigidity 45 $K_BT$ is simulated to investigate the response of gram-positive bacteria to nanopillared substrate as shown in Fig. \ref{fig:figuree6}. The nanopillars have varying heights, i.e. 50 nm, 100 nm, and 200 nm and a spacing of 170 nm. The simulation results indicate a markedly different response of bateria from those observed in scenarios involving bacteria with lower membrane bending rigidity. In contrast to simulations with low membrane bending rigidity, the bacterial membrane with high membrane bending rigidity in Fig. \ref{fig:figuree6} (a), (b), and (c) exhibited minimal deformation and bacteria was inclined to keep its geometrical shape intact. The bacteria can be observed to rest on top of nanopillars, where nanopillars effectively formed a mechanical support structure that ensured bacterial viability. This intriguing observation suggests that the height of nanopillars had a negligible impact on bactericidal effects on cylindrical bacteria  with higher membrane bending rigidity. The bacterial membrane tearing was not observed in cases with high membrane bending rigidity.  This finding underscore a significant contrast in the response of bacteria to nanopillars based on their membrane bending rigidity. While lower bending rigidity bacteria exhibited susceptibility to tearing mechanisms, their high bending rigidity bacteria displayed resilience against such bactericidal effects, irrespective of nanopillar height.  It needs to be pointed out that cylindrical bacteria can still be rendered nonviable if subjected to increased external forces. In this scenario, the nanopillars pierce or puncture the bacterial cell membrane, leading to the death of the bacteria.  It is worth noting that the killing of bacteria with high membrane rigidity can also occur with low nanopillar heights when subjected to increased external forces because the bactericidal mechanism in this case involves piercing or puncturing the bacteria's membrane. Consequently, it can be concluded that bacteria with high membrane bending rigidity are more resilient and challenging to kill compared to those with low bending rigidity.

Next, we investigated the influence of nanopillar spacing on cylindrical bacteria with varied membrane bending rigidity, i.e. 15 $K_BT$ vs 45 $K_BT$ as illustrated in Fig. \ref{fig:figuree7} and Fig. \ref{fig:figuree8}. With a nanopillar height of 200 nm in all cases, three distinct nanopillar spacing were considering, i.e. 170 nm, 240 nm, and 500 nm. As evidenced by Fig. \ref{fig:figuree7}, an increase in nanopillar spacing suggests a higher chance for bacteria to survive, particularly evident when examining Fig. \ref{fig:figuree7} (a) and (b) where low nanopillar spacing leads to the death of bacteria. From Fig. \ref{fig:figuree7} (b)  it can be observed that it is more likely for bacteria with low membrane bending rigidity to fit into the spaces between nanopillar without damage due to their capability to deform, while from Fig. \ref{fig:figuree8} (b) it can be observed that the bacteria with high membrane bending rigidity tends to keep their geometry and stay on top of nanopillar. When the nanopillar spacing increases beyond the diameter of the cylindrical bacteria, i.e. 500 nm, both type of bacteria easily fell between nanopillars and stay alive as shown in From Fig. \ref{fig:figuree7} and \ref{fig:figuree8}. These findings indicate that nanopillar spacing is relatively more critical in dictating bactericidal effects towards bacteria with low bending rigidity. This observation aligns with experimental results presented in Fig. \ref{fig:figureee1} (a), where, for nanopillar heights of 200 nm and 400 nm, reduced bactericidal effects were observed for nanopillar arrays with greater spacing between them for both high membrane bending rigidity bacteria and low membrane bending rigidity bacteria. Also, these computational findings supports that increasing nanopillar spacing would let us observe less bactericidal activity in gram-negative bacteria than in a gram positive bacteria i.e percentage decrease in non viable gram-negative bacteria would be higher than percentage decrease in non viable gram positive bacteria when nanopillar spacing is increased from experimental studies.

\subsection*{Spherical Bacteria}

\begin{figure}[hbt!]
\centering
\includegraphics[scale=0.6]{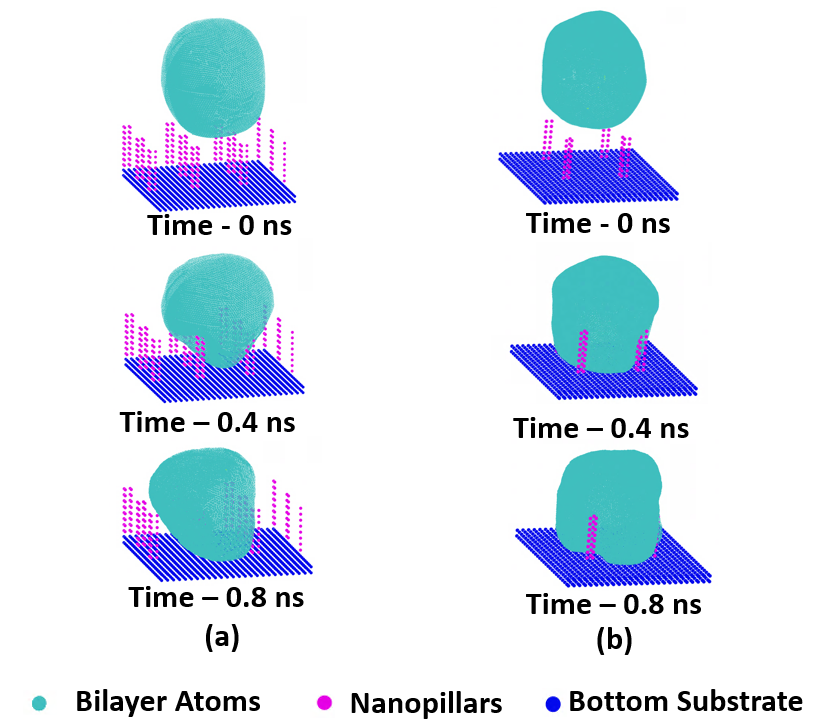}
\caption{ Effect of nanopillars spacing on spherical bacteria  with low bending rigidity membrane  (15 $K_bT$) (a) with nanopillar spacing = 170 nm (b) with nanopillar spacing = 240 nm. It can be noticed that with the progression of time and for all the cases spherical bacteria membrane deformed and the bacteria squeezed between the nanopillared arrays and reached the bottom substrate. It can be also noticed that with increase in the nanopillar spacing the bacteria membrane deforms less}
\label{fig:figuree11}
\end{figure}

We also examined the bactericidal effects of nanopillared surfaces on spherical bacteria with a diameter of 500 nm interacting with nanopillared surface with a nanopillar spacing of 170 nm and 240 nm. Similar to cylindrical bacteria, spherical bacteria with low membrane bending rigidity, i.e. 15 $K_BT$, can envelop the nanopillars with the height up to 200 nm while high membrane bending rigidity (45 $K_BT$) leads to minimal deformation of the bacterium and stay on top of nanopillars as shown in Fig. \ref{fig:figuree9} and \ref{fig:figuree10}. In Fig. \ref{fig:figuree9} (a) with short nanopillars of height 50 nm, the spherical bacterium sagged to entirely cover the nanopillars and descended to the bottom of the substrate without noticeable damage. With the height of the nanopillars increasing to 200 nm, the bacterium can be noted to partially touch the bottom but have part of the membrane suspended over the nanopillars. This is because the spherical bacteria is small enough to slide into the space between nanopillars. So, to be able to kill this bacteria we need to further reduce the nanopillar spacing.  We further investigated the effect of nanopillar spacing  on the bactericidal activity. With the increase in nanopillar spacing from 170 nm to 240 nm as shown in Fig. \ref{fig:figuree11}, spherical bacteria with low membrane bending rigidity can easily squeeze in between nanopillars and completely sink down to the bottom which creating a higher probability of  bacteria to survive. For high membrane bending rigidity bacteria as shown in Fig. \ref{fig:figuree10} and \ref{fig:figuree11}, the bacteria can stay on top of the nanopillars with all the varying heights and spacing used in this study due to higher membrane rigidity causing much less deformation in bacteria. Therefore, spherical bacteria with low bending rigidity, which is smaller than the cylindrical bacteria, are sensitive to both spacing and height of nanopillars. The nanopillars need to be densely packed and tall enough to have the bacteria cell wall suspended over the nanopillars, which can induce sagging and tearing of the membrane.

\begin{figure}[hbt!]
\centering
\includegraphics[scale=0.6]{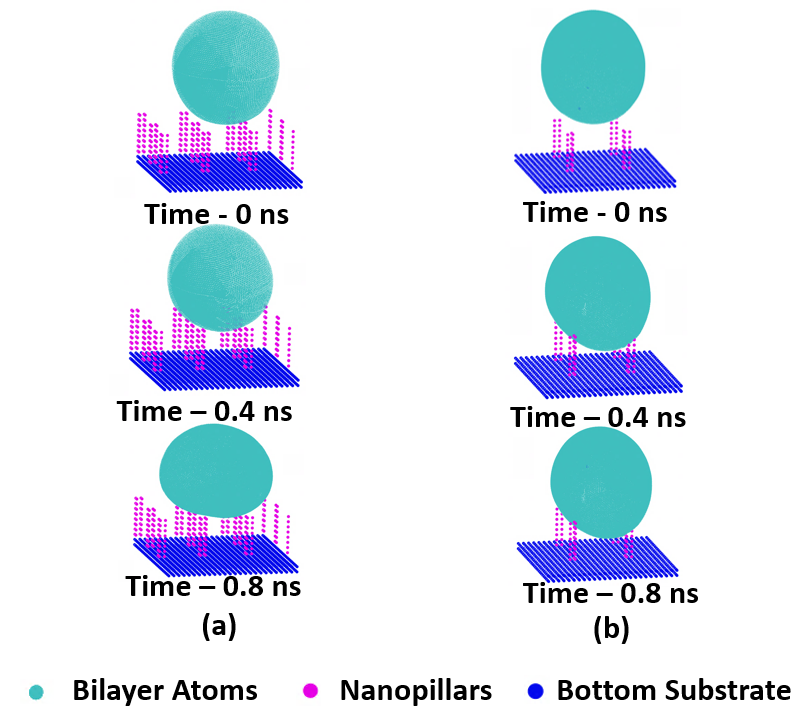}
\caption{ Effect of nanopillars spacing on spherical bacteria  with low bending rigidity membrane  (45 $K_bT$)(a) with nanopillar spacing = 170 nm (b) with nanopillar spacing = 240 nm. It can be noticed that with the progression of time and for all the cases spherical bacteria membrane shows less deformation and bacteria sits on top of nanopillar array}
\label{fig:figuree12}
\end{figure}

\noindent\textbf{Adhesion}

\begin{figure}[ht]
\centering
\includegraphics[scale=0.45]{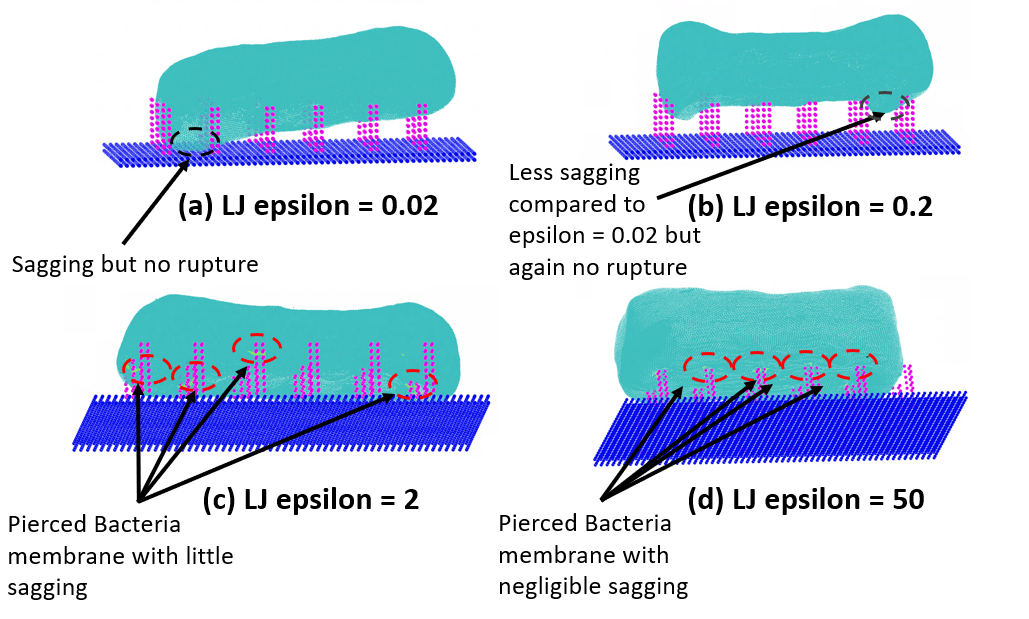}
\caption{Effect of changes in the $\epsilon$ parameter for the Lennard-Jones (LJ) interaction between the bacterial membrane and nanopillars:
(a) For $\epsilon = 0.02$, rupture is observed after 0.5 ns of interaction with the nanopillars; however, significant sagging of the bacterial membrane toward the substrate can be noted. (b) For $\epsilon = 0.2$, no rupture is observed after 0.5 ns of interaction with the nanopillars, but the sagging of the bacterial membrane is less pronounced compared to $\epsilon = 0.02$.
(c) For $\epsilon = 2$, and (d) for $\epsilon = 50$, piercing of the bacterial membrane is observed. Note: The bacteria and nanopillar model are being observed from the bottom, i.e., from the flat surface of the nanopillars.}
\label{fig:figure444}
\end{figure}

\begin{table}[hbt!]
\caption{\label{tab:tab3} Case studies by varying nanopillar height, bacteria's membrane bending rigidity and loading rate($\AA$). d=60 nm, s= 170 nm and h=height of nanopillars, B.R.= Bending rigidity of nanopillars in( $K_bT$ ) and L.R.= Loading rate on bacteria (eV. mole)/(\AA .grams)   }
\centering
\begin{tabular}{lcccc}
\hline
S.N0 & h   & B.R & L.R. & Results \\
 \hline
1 & 500 & 8.2 & 0.0002 & deformed but no damage \\
2 & 1000 & 8.2 & 0.0002 & deformed but no damage \\
3 & 2000 & 8.2 & 0.0002 & stretching and tearing \\
 \hline
4 & 500 & 45.4 & 0.0002 & little deformation \\
5 & 1000 & 45.4 & 0.0002 & little deformation \\
6 & 2000 & 45.4 & 0.0002 & little deformation\\
 \hline
7 & 2000 & 45.4 & 0.001 & pierced membrane \\

\hline
\end{tabular}
\end{table}

In this section, we studied the effect of changes in the interaction between nanopillars and bacteria, simulated using the Lennard-Jones (LJ) potential parameter $\epsilon$, on the bacterium-killing mechanism on nanopillared surfaces. The LJ potential mimics a portion of the attractive forces between the bacterium and the nanopillars, with the remaining attractive forces simulated by applying a uniform external force to the bacterium, referred to as the loading rate. We used a baseline case with $\epsilon = 0.02$, which represents the depth of the potential energy well in LJ potentials and approximately equates to $\Delta G_{iwi} = -0.0093 , mJ , m^{-2}$. Three additional cases were simulated by increasing $\epsilon$ to 0.2, 2, and 50, respectively, for 0.5 ns, with a uniform loading rate of $0.0002 , \text{eV}/(m \times \text{\AA})$ per CG atom of the bacterium model in all simulations.
From Fig. \ref{fig:figure444} (a) and (b), it can be observed that with $\epsilon = 0.02$ and $\epsilon = 0.2$, the bacterium membrane sags between the nanopillars, eventually tearing later in the simulation due to sagging. The tearing of the bacterium membrane occurs slightly later in case (a) with $\epsilon = 0.02$ compared to case (b) with $\epsilon = 0.2$. The sagging in both cases, shown in Fig. \ref{fig:figure444} (a) and (b), is more prominent near the bacteria and the nanopillar tip/edge, although no piercing or tearing of the bacterium membrane was observed within 0.5 ns of the simulation. It is also important to note that the sagging of the bacterium membrane near the cylindrical tip of the bacteria is less prominent in Fig. \ref{fig:figure444} (b) with $\epsilon = 0.2$ compared to Fig. \ref{fig:figure444} (a) with $\epsilon = 0.02$. This suggests that in Fig. \ref{fig:figure444} (a), the bacteria slightly attach at the nanopillar tip and then attempt to slip between the nanopillars, effectively transferring mass to one side. In contrast, in Fig. \ref{fig:figure444} (b), the bacteria remain stuck at the point of contact with the nanopillars due to stronger adhesion, resulting in slightly less sagging. When the interaction parameter $\epsilon$ is increased to 2 and 50, as shown in Fig. \ref{fig:figure444} (c) and (d), respectively, the bacterium membrane firmly attaches to the nanopillar at the point of contact and gets punctured or pierced at the same point within 0.5 ns. In these two cases, the bacteria cannot slide or move over the nanopillars as they did in cases (a) and (b), due to the very high attractive forces at the point of contact, likely simulating unrealistic conditions. These observations suggest that increasing the LJ parameter $\epsilon$ can be used to simulate different types of nanopillars made of various materials, as well as different geometric conditions with varying attractive forces with the bacterium membrane. This also indicates that the bactericidal failure mechanism of the membrane depends on the attractive force between the bacterium and the nanopillar. As the attractive forces increase, the bacterium-killing mechanism shifts from membrane tearing to membrane puncturing. Therefore, to simulate a realistic nanopillar-bacterium model, we used an LJ parameter of $\epsilon = 0.02$ in all simulations, with a loading rate of $0.0002 , \text{eV}/(m \times \text{\AA})$ per CG atom of the bacterium model.

\subsection*{Bactericidal mechanisms}

As we have previously discussed, most MD simulation based studies from previous literature have not investigated the dynamic activity of bacteria in the presence of nanopillared arrays. In this study, we have simulated the dynamic activity of bacterial membrane with accurate membrane bending rigidity for gram-positive/gram-negative bacteria using MD simulations for both gram-negative and gram-positive bacteria i.e. with low membrane bending rigidity (15 $K_BT$) and high membrane bending rigidity (45 $K_BT$) respectively. The bacterium  is usually suspended far away from the nanopillared surface such that LJ interaction force between nanopillared surface and bacterium membrane is close to zero. To simulate the attractive force between bacterial membrane and nanopillared surface we apply additional force to the bacteria (referred as the loading rate in this study), in addition to LJ interaction between membrane and nanopillars, which simulates the actual behavior of bacteria near the nanopillared surfaces. Loading rate helps simulate the bacterium membrane deformation and attachment when in contact with nanopillared surface in our MD simulations.  For gram-negative bacteria, when the bacterial membrane comes in contact with the nanopillared surface as shown in Fig. \ref{fig:figuree5} (b - Middle figure), the bacterial membrane starts deforming and form localized dents in the bacterial membrane, creating multiple ripples in the bacterial membrane near the point of contact with nanopillared arrays. As the time progresses, the bacterium tries to slide in between the nanopillared surfaces creating uneven distribution of mass by moving some of its internal fluids towards the edge/tip of bacterium. During this phase, the bacterium membrane stretches in the middle portion, eliminating ripples (in the middle portion), as both the ends of cylindrical bacterium try to glide in between nanopillars as shown in Fig. \ref{fig:figuree5} (c - Bottom figure). If the nanopillars are not long enough and the nanopillar spacing is high enough, the bacterium tip will reach at the bottom of nanopillars and will gets support for the overhanging bacterium from the bottom flat surface and thus the bacterium survives, as shown in Fig. \ref{fig:figuree5} (b,c - Bottom figure). But if the nanopillars are long enough and the nanopillar spacing is high enough, the hanging weight of bacterium, near the ends of cylindrical bacterium, between the nanopillars stretches the bacterium membrane so much that it tears near the point of contact with one of the nanopillars (likely the point of maximum stress), as shown in Fig. \ref{fig:figuree5} (c - Bottom figure). Note that, this bactericidal mechanism is different from  Pogodin et. al \cite{Pogodin2013_10} which suggested that at all points of contact between nanopillared array and bacterial membrane the bacterium membrane sags in between nanopillared arrays uniformly and tears near the center of sagged portion of bacterium membrane between two nanopillars whereas our bactericidal mechanism suggests that bacterium does not sag between all the nanopillars, but at few nanopillars near the edge of the cylindrical bacteria and stretches the bacterial membrane so much that it tears (tearing happens very close to the point of contact with nanopillar near the sliding tip of bacterium). Similarly, MD study of Salatto et. al \cite{Salatto2023_Structure}  suggested that increasing adhesion between nanopillars and bacterial membrane cause bacterium membrane to take the shape of nanopillars thereby increasing stress concentration at the point of contact between nanopillars and bacterium membrane. Since, we are studying dynamic response of bacterium in the presence of  nanopillared surfaces, therefore our study suggests that bacterium membrane might exhibit higher membrane deformation near the edge of bacterium where its suspended weight might also be a factor in inducing stretching and tearing of bacterium membrane apart from the adhesive interaction between bacterium and nanopillars. Similarly, when gram-positive bacteria comes in contact with the nanopillared surface, as shown in Fig. \ref{fig:figuree6} (a,b,c - Middle figure), the bacterial membrane starts deforming uniformly but does not form localized dents in the bacterial membrane. Since, the bacterial membrane bending rigidity is very high the nanopillars acts as a support structure and helps suspend the bacterium over nanopillared surface. As the time progresses, the bacterium membrane deforms a little more till the flexural stiffness of the curved bacterium is able to counter the loading rate (force) applied on the bacterium. At this point the membrane deformation stops. Therefore the gram-positive bacterium survives over a similar nanopillared surface on which a gram-negative bacterium got killed, as shown in Fig. \ref{fig:figuree6} (a,b,c - Bottom figure). In the case of gram-positive bacteria the bacterium deforms as a whole and no localized dents are observed on the membrane near the point of contact with nanopillars. For killing a gram-positive bacterium we need to increase the loading rate, as shown in Fig. \ref{fig:figuree16}, where the piercing mechanism activates. During piercing the bacterium membrane gets punctured at the point of contact between nanopillar and membrane when the loading rate is increased (force) on the bacterium. Since, the bending rigidity of bacterium membrane is already high no localized stretching in bacterium membrane is observed.

\begin{figure}[hbt!]
\centering
\includegraphics[scale=0.55]{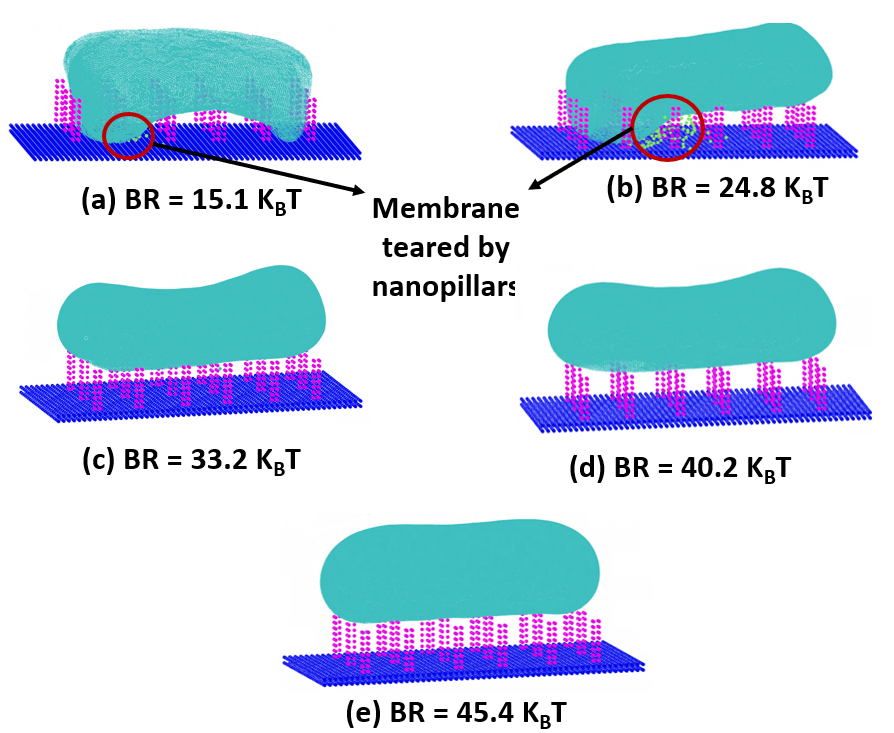}
\caption{ Effect of change in bacterial membrane's bending rigidity on cylindrical bacteria (a) with bending rigidity = 15.1 $K_bT$ (b) with bending rigidity = 24.8 $K_bT$ 
(c) with bending rigidity = 33.2 $K_bT$ (d) with bending rigidity = 40.2 $K_bT$ (e) with bending rigidity = 45.4 $K_bT$. It can be noticed that with the for cases (a) and (b) we can observe bacteria's membrane stretching and getting teared near point of contact between nanopillar and bacteria membrane but for cases (c),(d) and (e) the bacteria sits on top of nanopillars and survives. All the results are after 1 ns of simulation}
\label{fig:figuree13}
\end{figure}

\begin{figure}[hbt!]
\centering
\includegraphics[scale=0.6]{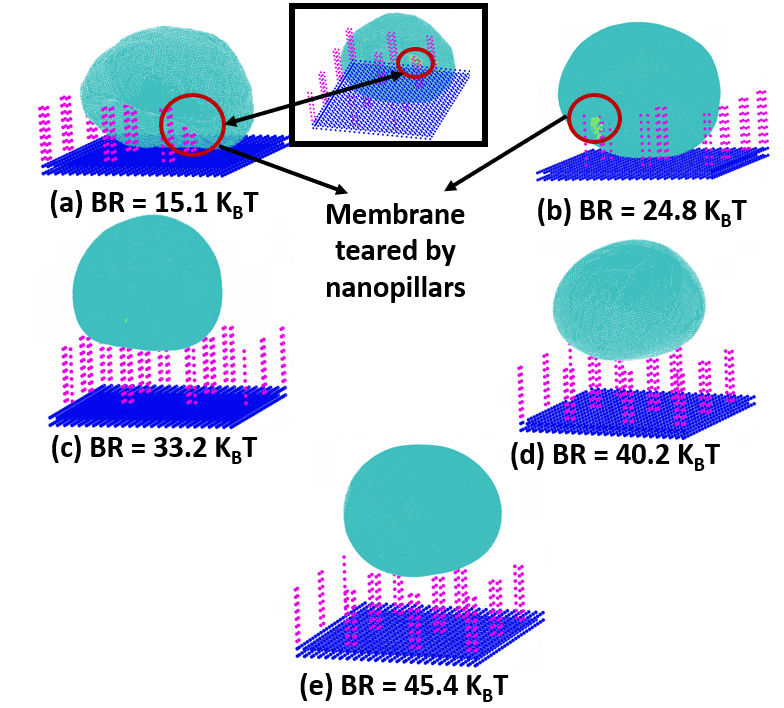}
\caption{ Effect of change in bacterial membrane's bending rigidity on spherical bacteria (a) with bending rigidity = 15.1 $K_bT$ (b) with bending rigidity = 24.8 $K_bT$ 
(c) with bending rigidity = 33.2 $K_bT$ (d) with bending rigidity = 40.2 $K_bT$ (e) with bending rigidity = 45.4 $K_bT$. It can be noticed that with the for cases (a) and (b) we can observe bacteria's membrane stretching and getting teared near point of contact between nanopillar and bacteria membrane but for cases (c),(d) and (e) the bacteria sits on top of nanopillars and survives. All the results are after 1 ns of simulation}
\label{fig:figuree14}
\end{figure}

It has been observed that the bacterial membrane bending rigidity can largely govern the deformation behaviors of bacteria on top of nanopillared surfaces. Therefore, we gradually varied the membrane bending rigidity of bacteria from 15.1 to 45.4 $K_bT$ as shown in Fig. \ref{fig:figuree13} and \ref{fig:figuree14} to observe the influence of change in bending rigidity of bacterial membrane on bactericidal mechanism. The nanopillars in all the cases were 200 nm tall and 170 nm spaced. For both cylindrical and spherical bacteria, the results clearly show that as the bending rigidity of the bacteria membrane increases, the deformation in the bacteria membrane gradually decreases with less sagging between the nanopillars which leads to less stretching in the membrane induced by suspension between nanopillars. When the membrane bending rigidity is smaller than 33.2 $K_bT$, the bacteria can sag between nanopillars and have the membrane suspended and stretched, which may potentially lead to failure of bacteria membrane. When the membrane bending rigidity becomes larger than 33.2 $K_bT$, the bacteria can deform less accordingly when the come in contact with nanopillared tip. High bending rigidity bacterium can keep their geometry intact (stay in their original shape i.e. either cylindrical shape  or spherical shape) observing much less membrane deformation than the bacterium with low membrane bending rigidity. This can allow them comfortably sit on top of the nanopillars with minimal contact with the nanopillars. 

\begin{figure}[hbt!]
\centering
\includegraphics[scale=0.5]{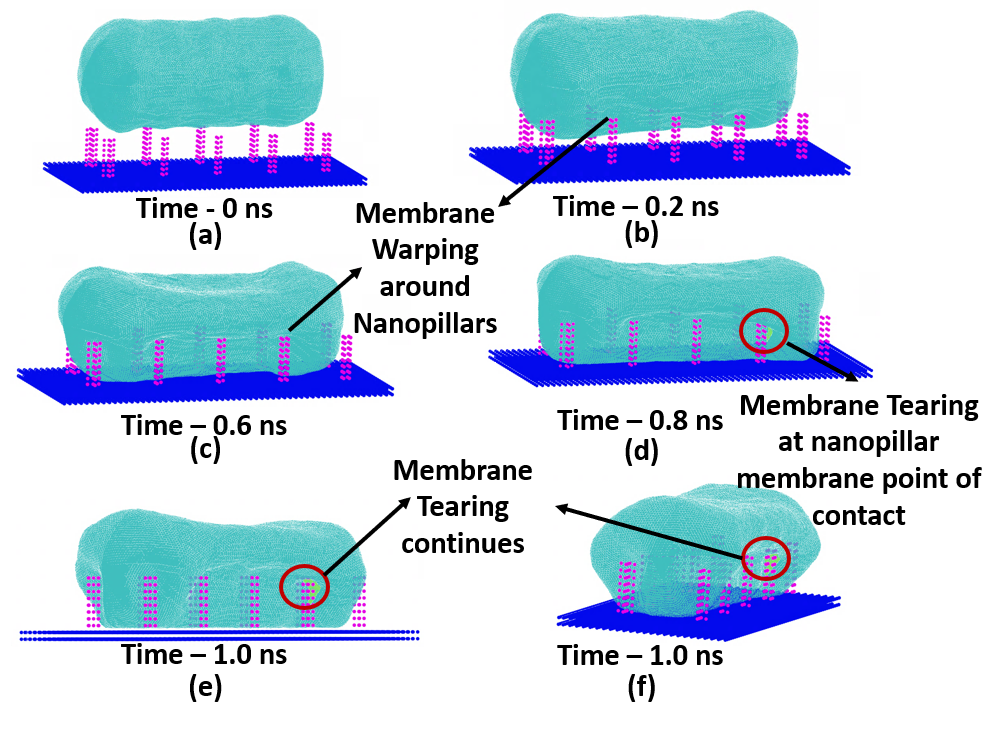}
\caption{ Tearing failure mechanism (a) at time = 0 ns : No contact with nanopillars (b) at time = 0.2 ns : Localized deformation in bacteria's membrane at point of contact (c) at time = 0.6 ns : Bacteria's membrane warping around nanopillars as it is sinking down due to gravity (d) at time = 0.8 ns : Bacteria's membrane stretched and teared at the point of contact between nanopillar and membrane (e) and (f) at time = 1.0 ns : Bacteria's membrane tearing continues shown from two different angles}
\label{fig:figuree15}
\end{figure}

\begin{figure}[hbt!]
\centering
\includegraphics[scale=0.55]{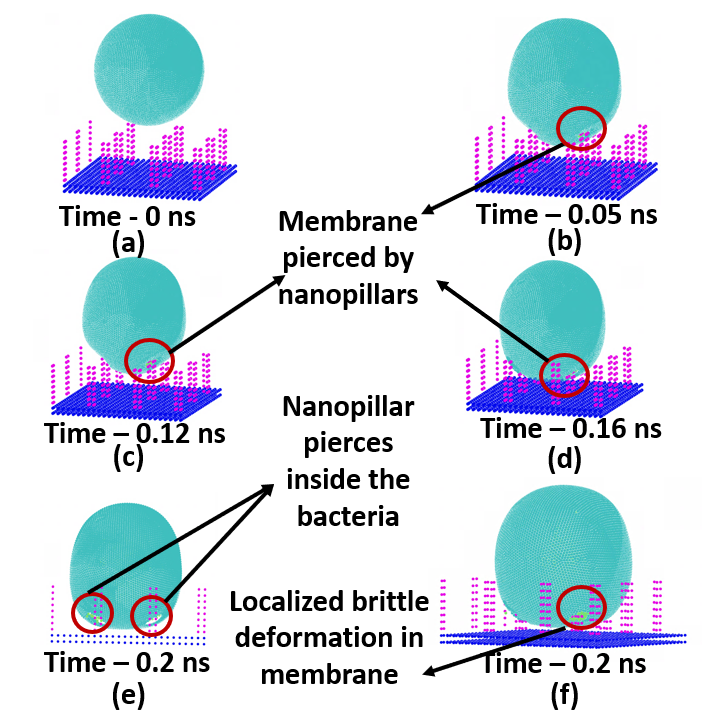}
\caption{ Piercing failure mechanism (a) at time = 0 ns : No contact with nanopillars (b) at time = 0.05 ns : Contact with nanopillars and puncture initiated at the point of contact (c) at time = 0.12 ns : Nanopillar pushes inside the bacteria (d) at time = 0.16 ns : Nanopillar further pushes inside the bacteria  (e) at time = 0.2 ns : Full nanopillar pierces inside the bacteria. Note: The bacteria still retains the spherical shape }
\label{fig:figuree16}
\end{figure}

Therefore, it can be concluded that when comparing the geometry, i.e. cylinder vs sphere, the membrane bending rigidity plays a more critical role on the failure of membrane and therefore the bactericidal mechanism of nanopillared surface/substrate. For gram-negative bacteria with low membrane bending rigidity, the failure mode is stretching and tearing of bacteria's membrane due to sagging induced by the bacteria/nanopillar interaction and environmental loading. Tearing happens when the bacteria membrane is stretched beyond a certain limit while sagging to reach the bottom surface of the nanopillared substrate as can be seen from  Fig. \ref{fig:figuree15}. With short nanopillars as shown in Fig. \ref{fig:figuree5} (a) $\&$ (b) and \ref{fig:figuree9} (a) $\&$ (b), the bacteria membrane reach the bottom for extra support before tearing happens. This suggests a critical height of nanopillars to induce significant sagging of bacteria membrane to achieve the bactericidal effect. As for the gram-positive bacteria with high membrane bending rigidity, failure is more likely to happen as piercing and puncturing where the membrane is in contact with the nanopillars. Therefore, the height of nanopillars doesn't affect the bactericidal effect. For gram positive bacteria bactericidal effect can be more impacted by the interaction between nanopillars and bacteria as shown in Fig. \ref{fig:figuree16}.

\subsection*{Experimental Validation}

\begin{figure*}[hbt!]
\centering
\includegraphics[scale=0.72]{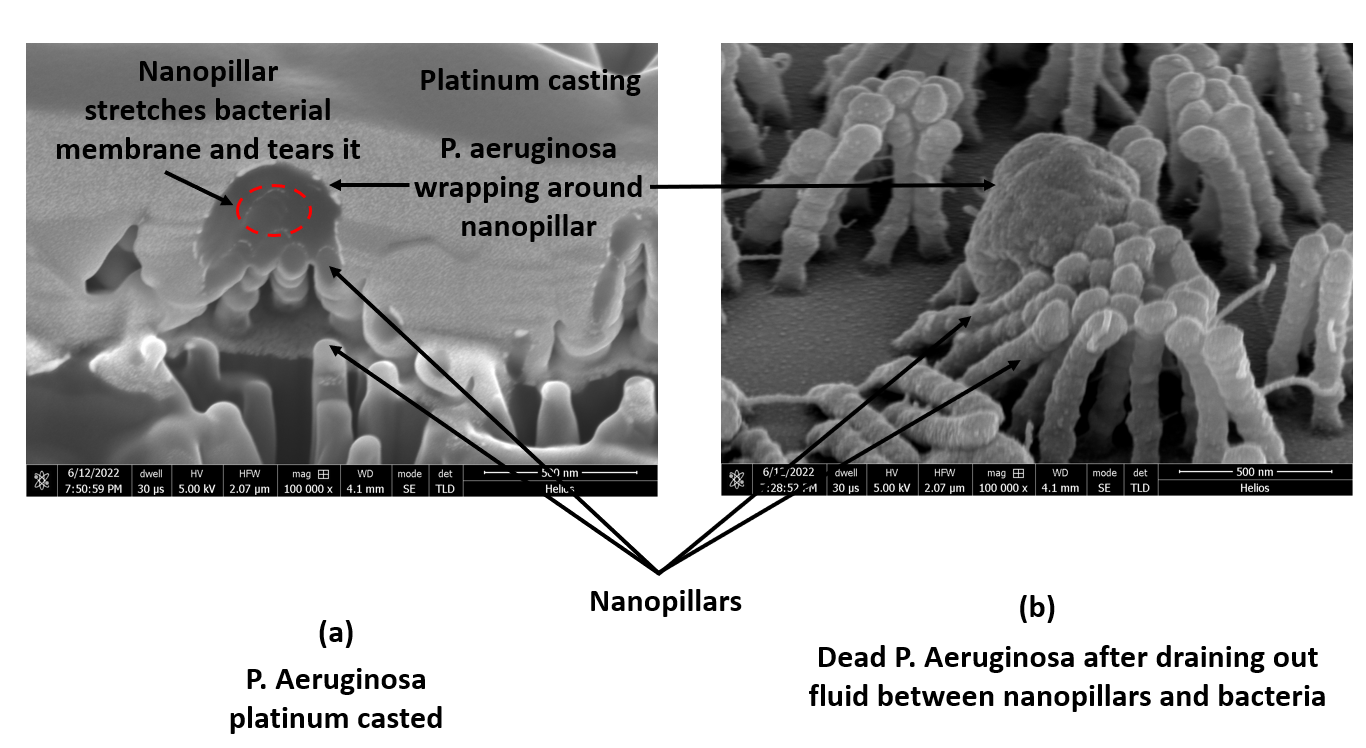}
\caption{Figure showing bactericidal effects of 800 nm long nanopillars substrate for gram-negative bacteria P. aeruginosa (a) shows a platinum casted P. aeruginosa on a nanopillared substrate where P. aeruginosa is  wrapping around multiple nanopillars. It is important to note that P. aeruginosa membrane undergoes large deformation and we can also observe that one of the nanopillar stretches the bacterial membrane so much that it tears the membrane and goes inside the cytoplasm/inner fluid of P. aeruginosa (encircled with red)  (b) shows a dead  P. aeruginosa wrapping around multiple nanopillars. This image is captured after the fluid between nanopillared substrate and bacteria is drained out. The warping of the bacteria under nanopillared surface is obvious and we can also observe that some nanopillars might have teared the P. aeruginosa's membrane and are inside the bacterium.  Note: The nanopillars coalesce after the fluid is drained out after the incubation process but during the incubation process the nanopillars were straight up (\cite{Yi2023_Asmart} Figure 3 (b) from Yi et. al.)
}
\label{fig:figuree17}
\end{figure*}

\begin{figure*}[hbt!]
\centering
\includegraphics[scale=0.72]{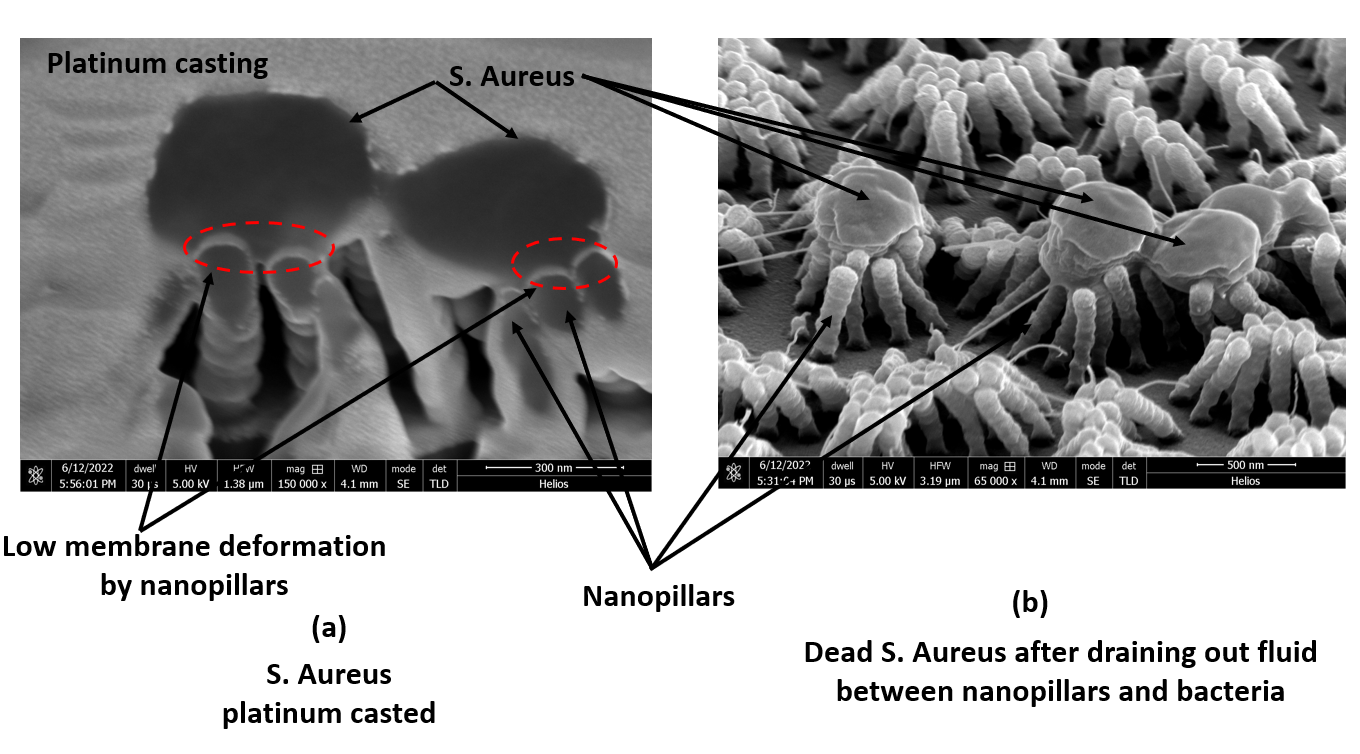}
\caption{Figure showing bactericidal effects of 800 nm long nanopillars substrate for gram-positive bacteria S. aureus (a) shows a platinum casted S. aureus on a nanopillared substrate where S. aureus sits on top of multiple nanopillars without significant deformation in bacterial membrane (encircled with red)  (b) top view of a few dead S. aureus bacteria sitting on top of nanopillared substrate. All the three S. aureus are presumed to be dead because we can observe a loss of turgor pressure in all the three S. aureus shown and most likely these bacterium were punctured/pierced during the incubation process. Note: The nanopillars coalesce after the fluid is drained out after the incubation process but during the incubation process the nanopillars were straight up (\cite{Yi2023_Asmart} Figure 3 (b) from Yi et. al.)
}
\label{fig:figuree18}
\end{figure*}

\begin{figure*}[hbt!]
\centering
\includegraphics[scale=0.75]{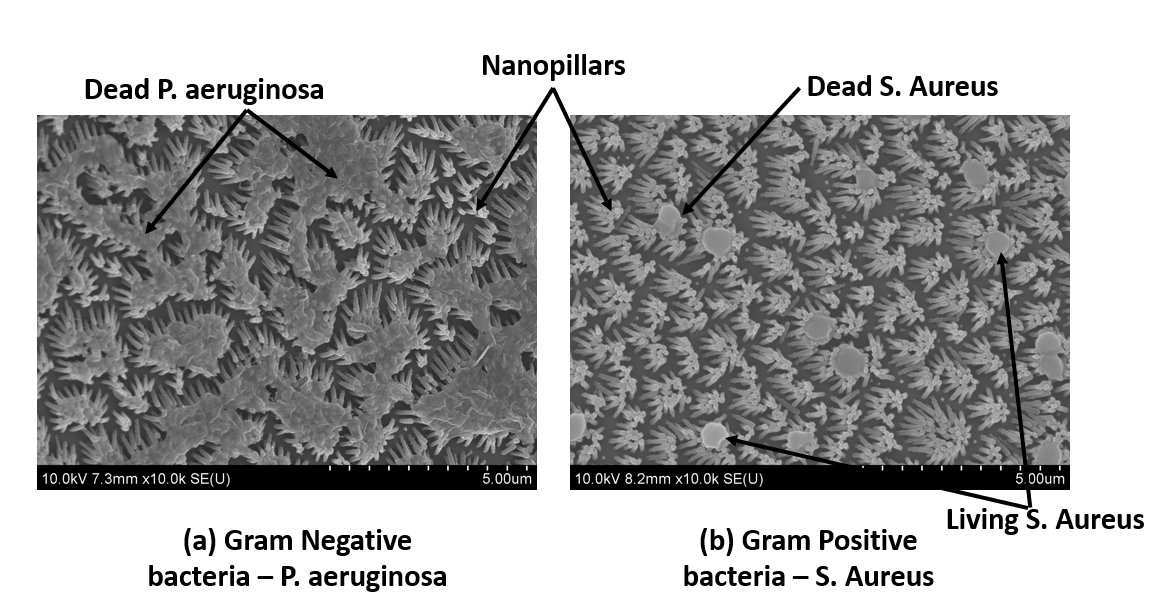}
\caption{Figure showing effect of bacterial membrane bending rigidity on bactericidal effects of 800 nm long nanopillars substrate  (a) shows lots of dead P.aeruginosa after removal of fluid (b) shows many surviving S.aureus after removal of fluid. his suggests that even for same conditions we observe more deaths of P.aeruginosa compared to S.aureus due to higher bending rigidity of S.aureus.
}
\label{fig:figuree20}
\end{figure*}

The bactericidal effects of nanopillared substrate on gram-negative bacteria P. aeruginosa and gram positive bacteria S. aureus were observed after letting the nanopillared substrate and bacterium to  interact for 30 minutes as shown in Fig. \ref{fig:figuree17} and \ref{fig:figuree18}. Note that we are showing the results of P. aeruginosa and not E. coli as they have similar membrane bending rigidity therefore similar results were expected. The size of nanopillars used in this experimental study was  800 nm in height with 50 nm diameter. Fig. \ref{fig:figuree17} (a) shows a platinum casted cross-section of nanopillared substrate with P.aeruginosa being wrapped around nanopillar. Casting is done to preserve the shape of the bacteria and nanopillars at the point of casting so that we can observe it later under scanning electron microscope. A few nanopillars can be seen possibly penetrate through the bacteria in Fig. \ref{fig:figuree17} (a) shows a platinum casted P. aeruginosa on a nanopillared substrate where P. aeruginosa is wrapping around multiple nanopillars.
It is important to note that P. aeruginosa membrane undergoes large deformation and we can also observe that one of the nanopillar stretches the bacterial membrane so much that it tears the membrane and goes inside the cytoplasm/inner fluid of P. aeruginosa (encircled with red). Fig. \ref{fig:figuree17} (b) shows the image of P.aeruginosa with nanopillared surface after draining out of all the liquid to be able to observe under scanning electron microscope. Here, we can observe that P.aeruginosa is trapped between multiple nanopillars by wrapping around them. The bacterium shows lost turgor pressure and is possibly dead.  The warping of the bacteria under nanopillared surface is obvious and we can also observe that some nanopillars might have teared the P. aeruginosa’s membrane and are inside the bacterium.  Note that the nanopillars clustering is explained by Yi et. al. \cite{Yi2023_Asmart} by cell fixation protocol which is beyond the scope of this study. Although, the nanopillars coalesce after the fluid is drained out after the incubation process but during the incubation process the nanopillars were straight up ((\cite{Yi2023_Asmart}) Figure 3 (b) from Yi et. al.).  Similarly, Fig. \ref{fig:figuree18} (a) shows a platinum casted cross-section of nanopillared substrate with S. aureus being sitting on top of nanopillared substrate. The figure shows two spherical S. aureus sitting on top of nanopillars. As the bacteria can be seen sitting on top of nanopillared surface with little to no sign of membrane being curved around the nanopillars. Thus, it can be concluded that this bacteria would have most likely stayed alive if not casted. Fig. \ref{fig:figuree18} (b) few dead S. aureus bacteria sitting on top of nanopillared substrate. All the three S. aureus are presumed dead because we can observe a loss of turgor pressure in all the three S. aureus shown and most likely these bacterium were punctured/pierced during the incubation process. This can be further validated by Fig. \ref{fig:figuree20} where we observe that after draining the fluid for the same nanopillar configuration and area we can observe higher fatality rate of P.aeruginosa compared to S.aureus. Fig. \ref{fig:figuree20} (a) suggests that almost all P.aeruginosa are dead on 800 nm nanopillared array whereas from Fig. \ref{fig:figuree20} (b) we can observe that a few S.aureus are dead but mostly all the S.aureus are surviving.

This validates our computational simulations finding from Fig. \ref{fig:figuree13} and Fig. \ref{fig:figuree14} where we suggested that bactericidal activity for nanopillars is dependent on the bending rigidity of the membrane and for all the gram-negative bacteria the prominent bactericidal mechanism should be tearing and stretching of the membrane. From  Fig. \ref{fig:figuree17} (a) it can also be noticed that  P.aeruginosa  warped around nanopillars and gets killed possibly by membrane stretching and tearing near nanopillar tip as suggested by our computational study. From Fig. \ref{fig:figuree18}  it can be noticed that for gram positive bacteria S. aureus was able to survive the 800 nm nanopillars. This further validates our findings from Fig. \ref{fig:figuree13} and Fig. \ref{fig:figuree14} where we suggested that bacteria with high membrane bending rigidity does not get killed by nanopillared surface even with long nanopillar. Also, as the membrane bending rigidity increases the bacteria becomes less susceptible to failure in the presence of long nanopillars. The bactericidal effects in high membrane bending rigidity bacteria can only be increased by increasing the bacteria's impact (loading rate) with nanopillars i.e. by increasing the impact force between bacteria and nanopillars and in only such cases membrane piercing can be observed  as shown in Fig. \ref{fig:figuree15}.

\section*{Conclusion}
This article presents the development of a coarse-grained (CG) model for conducting molecular dynamics (MD) simulations, aimed at exploring the interaction between bacterial membranes and nanopillar arrays. As per our knowledge, till date no coarse grained model has been developed which captures the physical/mechanical properties of bacteria's membrane specifically bending rigidity of bacterial membrane to investigate the bactericidal effects on nanopillared surfaces. Our coarse grained model successfully captures the bending rigidity of three layered bacteria's membrane i.e inner membrane, peptidoglycan layer and outer membrane with a single layer coarse grained model. Previous studies have focused on capturing the physical and chemical properties which do not allow for broader coarse graining of bacterial membrane as in these studies the focus was to coarse grain each of the three bacteria's layer separately which makes the bacterial membrane model very bulky and computationally expensive. Bending rigidity is one key feature of bacterial membrane which plays an important role in bacteria's survival when encountered with nanopillared surfaces. Both gram-positive and gram-negative bacteria usually show a varied range of bacterial membrane bending rigidity i.e 13 $\pm$ 5  K$_B$T to 43 $\pm$ 5 K$_B$T where the former has much rigid bacterial membrane with a large value of bending rigidity and the latter has a much flexible bacterial membrane with a low value of membrane bending rigidity. Keeping this in mind we have successfully created a bacteria model using a fluidmembrane pair potential \cite{FU2017_21} in LAMMPS where we can alter its membrane's bending rigidity by altering its parameters defined in the input script in LAMMPS. This eliminated the need of creating different models for simulating bacteria with different membrane bending rigidity.

This study also highlights the potential of coarse grained MD simulations (CG-MD) in unraveling the underlying physical mechanisms of bio-inspired nanostructured antifouling coatings/nanopillared surfaces. We studied the impact of nanopillar height, spacing and bending rigidity of bacterial cell deformation on nanopillared surfaces to investigate its potential killing mechanisms. Note that unless specified all the discussed results are valid for cylindrical bacteria with low loading rates. The findings of this study indicates that increasing nanopillar height increases bactericidal effects in bacteria with low membrane bending rigidity. This study also reveals that, for bacteria with low membrane bending rigidity, there exists a critical nanopillar height necessary to induce stretching and tearing which can lead to the subsequent bacterial death. Longer nanopillars with heights greater than or equal to 200 nm can stretch the membrane of bacteria with low bending rigidity, eventually tearing the membrane and leading to bacterial death. Conversely, in case of bacteria with high bending rigidity, increasing nanopillar height has no significant bactericidal effect and bacterium survived irrespective of nanopillar height by sitting on top of nanopillared surfaces (without much deformation). Also, for bacteria with higher membrane bending rigidity, no such critical height was observed above which we can observe higher bactericidal effects. However, increasing the loading rate allows for the piercing or puncturing of the high membrane bending rigidity bacterial cell walls, resulting in bacterial death. Thus, we can kill a high membrane bending rigidity bacteria through piercing/puncturing with high loading rate making these high bending rigidity bacteria tougher to kill than low membrane bending rigidity bacteria. 
Similarly, increasing nanopillar spacing, decreases bactericidal efficiency for low membrane bending rigidity bacteria as the bacteria squeezes and slides between the nanopillared surfaces to survive. On the other hand, high membrane bending rigidity bacteria survives irrespective of the nanopillar spacing. We also investigated the shape effects of bactericidal activity for smaller spherical bacteria of diameter 500 nm. The only difference we found that increasing nanopillar spacing decreases the bactericidal effects in spherical bacteria more than cylindrical bacteria as the small bacteria can squeeze between nanopillars easily and effectively and can survive. We also investigated the effect of change in bending rigidity of bacteria's membrane and can conclude that as the membrane bending rigidity of bacteria increases it makes less susceptible to membrane stretching/tearing and their survival likelihood increases. This suggests that interaction between nanopillar and bacterial membrane is more critical for gram positive bacteria where geometry of nanopillar, material of nanopillar a point of contact etc. may play a significant role in the bactericidal activity as these are the factors that can be altered to change the interaction between nanopillar and bacterium membrane. Also, for gram negative bacteria with lower membrane bending rigidity the  configuration of nanopillared surfaces i.e. length, spacing, geometry etc. creates a larger impact on bactericidal mechanisms.

In all this study concludes that we can have two distinct bactericidal mechanisms i.e tearing associated with bacteria with low membrane bending rigidity and piercing/puncturing associated with bacteria having high membrane bending rigidity which is only activated at high loading rates. This suggests that piercing is less frequent as high loading rate is synonymous to high impact which has a less likelihood in real world scenario. Thus, puncturing should be a less prominent bactericidal mechanism than tearing and piercing. Therefore nanopillared antifouling surfaces are more effective against gram-negative bacteria than gram-positive.   These insights provide a fundamental understanding of the bactericidal effects of bio-inspired nanostructured surfaces and pave the way for the design of optimized antifouling surfaces for diverse medical applications.

\section*{Acknowledgments}

The authors acknowledge the support provided by University of Illinois at Urbana Champaign and NSF support under Grant No. 2015292. All MD simulations were done by using LAMMPS, an open source code distributed by Sandia National Laboratories

\section{Appendix}
\noindent \textbf {Experimental Method}

\noindent \textbf {Preparation of the polyimide nanopillared surface} Silicon dioxide (SiO$_2$) 90 nm thick was grown on a silicon wafer using thermal oxidation. To make the SiO$_2$ surface hydrophilic, oxygen plasma treatment was applied using an 18-watt Harrick Plasma cleaner at 500 millitorr for 10 minutes. Next, a solution containing 5\% by weight of monodisperse polystyrene nanospheres (Thermo Fisher Scientific) was prepared in a 400:1 mixture of isopropyl alcohol and Triton X-100 (Sigma-Aldrich), which was then spin-coated onto the SiO$_2$ surface at 6000 rpm for 30 seconds. This process resulted in the formation of a hexagonally close-packed monolayer of nanospheres. The nanospheres were then selectively etched in oxygen plasma using a Plasma-Therm system operating at 100 watts radio-frequency power with an oxygen flow rate of 20 standard cubic centimeters per minute (SCCM) to maintain a chamber pressure of 100 millitorr. The etching rate was approximately 100 nanometers per minute. A thin layer of chromium (Cr) approximately 10 nanometers thick was uniformly deposited using electron-beam evaporation using a Temescal system. The nanospheres, with the chromium layer on top, were removed by dissolving the polystyrene in chloroform from Sigma-Aldrich, leaving behind a perforated chromium mask on the SiO$_2$ surface. The pattern from the chromium mask was then transferred to the underlying SiO$_2$ layer through etching in CF$_4$ plasma from a Plasma-Therm system, using 20 SCCM CF$_4$ at 50 millitorr and 300 watts. The chromium layer was subsequently removed using a chromium etchant (Transene), and deep-silicon reactive ion etching (RIE) was employed to create nanowell arrays as a template for micromolding. Each cycle of the deep RIE process, conducted with an inductively coupled plasma (ICP) system from SPTS Technologies, involved an etching step where a mixture of 130 SCCM SF$_6$ and 13 SCCM O$_2$ was used to achieve the desired etch characteristics. The fabrication process continued with plasma treatment using a coil power of 600 watts and a platen power of 12 watts for 5 seconds, followed by a deposition step where a flow of 85 standard cubic centimeters per minute  of C$_4$F$_8$ maintained a fluorocarbon plasma with a coil power of 600 watts and a platen power of 0 watts for 3 seconds. Each cycle of this process etched approximately 20 nanometers to create nearly vertical sidewalls. Hydrofluoric acid (HF) was then used to remove the SiO$_2$ mask, completing the fabrication of the templates. In the subsequent molding process, a solution of poly (pyromellitic dianhydride-co-4,4-oxydianiline) amic acid from Sigma-Aldrich was applied to the master template using doctor blading, controlling the thickness. The coated template was degassed in vacuum for 12 hours. Subsequently, a vacuum annealing process at 250°C under 1 torr for 1 hour cross-linked the polyamic acid liquid precursors, forming a solid polyimide film through thermal imidization. A slow ramping rate of 3° to 5°C per minute was crucial to encourage the diffusion of the polyamic acid precursors into the nanowells with nanometer-scale hydraulic radius. Once cooled to room temperature, the polyimide film, featuring high-density nanopillar arrays on one surface, was peeled off from the master template as a foil, retaining a flat top surface.

\noindent \textbf {Bactericidal Assessment of Nanostructured Polymer Films} To evaluate the bactericidal efficacy, viability assays were conducted using standard plate count methods. Initially, cultures of E. coli MG1655 and S. aureus USA 300 were incubated in Luria-Bertani (LB) solution. Subsequently, these cultures were suspended in 1x phosphate-buffered saline (PBS) and adjusted to an optical density (OD600) of 0.1. Next, 40 $\mu$L of the appropriately diluted bacterial solution was applied to a 1 cm² area of the test surface. The samples were then incubated for duration ranging from 1.5 to 3 hours. After the incubation period, the bacterial cells were removed from the surfaces through rinsing with 1 mL of PBS. The resulting bacterial suspensions underwent serial 1:10 dilutions. Each dilution was evenly spread onto LB agar plates and incubated at 37°C for 12 hours. The colonies that developed on these plates were subsequently counted, allowing for the determination of the number of colony-forming units per mL. It was assumed that the count of colony-forming units corresponded to the quantity of live cells in suspension.

\noindent \textbf {Scanning Electron Microscopy (SEM) Examination of Biofilm Formation} 

For SEM analysis, the samples underwent fixation using formaldehyde. Specifically, the samples were immersed in a 4\% (v/v) paraformaldehyde solution in PBS for 30 minutes at room temperature. Following this fixation step, the samples were rinsed with PBS and subjected to a dehydration process using a series of graded alcohol solutions, including 25\%, 50\%, 75\%, 90\%, and two rounds of 100\%, with each step lasting 15 minutes. Subsequently, all the samples were air-dried within a desiccator and then coated with a layer of gold utilizing a sputter coater. Finally, the samples were examined through scanning electron microscopy (SEM) using a Hitachi S4800 SEM instrument. To initiate biofilm formation by P. aeruginosa, the bacteria were first cultured in LB broth overnight at 37°C. After this incubation, they were washed three times and diluted to the desired concentrations, ranging from $10^4$ to $10^6$ cells per mL, in M63 minimal medium, which includes magnesium sulfate, glucose, and casamino acids. Subsequently, 100 $\mu$L of the prepared bacterial dilution was applied to a 1 cm² area of the test surface. The samples were then incubated in a humidified chamber, created using a Styrofoam box containing a pre-wetted paper towel, for a duration ranging from 24 to 72 hours.

% Uncomment if using bibtex (default)
\bibliography{sample}

% Uncomment if using biblatex
% \printbibliography

\end{document}